\documentclass[sigconf]{acmart}

\usepackage{caption}
\usepackage{subcaption}
\usepackage{tabularx}

\newcommand{\hl}[1]{{\color{blue}#1}}
 \renewcommand{\hl}[1]{{#1}}
\newenvironment{hl*}{\color{blue}}{}
\renewenvironment{hl*}{}{}

\newcommand{\todo}[1]{{\color{red}#1}}
 \renewcommand{\todo}[1]{{}}

\newcommand{\tpc}[1]{{\textbf{#1}}}
\renewcommand{\tpc}[1]{{#1}}

\newcommand{\study}[1]{{\color{green}#1}}
\renewcommand{\study}[1]{{#1}}

\AtBeginDocument{%
  \providecommand\BibTeX{{%
    \normalfont B\kern-0.5em{\scshape i\kern-0.25em b}\kern-0.8em\TeX}}}

\copyrightyear{2022} 
\acmYear{2022} 
\setcopyright{rightsretained} 
\acmConference[CHI '22]{CHI Conference on Human Factors in Computing Systems}{April 29-May 5, 2022}{New Orleans, LA, USA}
\acmBooktitle{CHI Conference on Human Factors in Computing Systems (CHI '22), April 29-May 5, 2022, New Orleans, LA, USA}
\acmDOI{10.1145/3491102.3517733}
\acmISBN{978-1-4503-9157-3/22/04}

\begin{document}

\title{Adorned in Memes: Exploring the Adoption of Social Wearables in Nordic Student Culture}

\author{Felix Anand Epp}
\email{mail@felix.science}
\orcid{0000-0001-6252-7244}
\affiliation{%
  \institution{Aalto University}
  \city{Espoo}
  \country{Finland}
}

\author{Anna Kantosalo}
\email{anna.kantosalo@helsinki.fi}
\orcid{0000-0001-7458-4297}
\affiliation{%
  \institution{Aalto University}
  \city{Espoo}
  \country{Finland}
}
\affiliation{%
  \institution{University of Helsinki}
  \city{Helsinki}
  \country{Finland}
}

\author{Nehal Jain}
\email{jainehal03@gmail.com}
\affiliation{%
  \institution{Aalto University}
  \city{Espoo}
  \country{Finland}
}

\author{Andrés Lucero}
\email{lucero@acm.org}
\affiliation{%
  \institution{Aalto University}
  \city{Espoo}
  \country{Finland}
}

\author{Elisa D. Mekler}
\email{elisa.mekler@aalto.fi}
\orcid{0000-0003-0076-6703}
\affiliation{%
  \institution{Aalto University}
  \city{Espoo}
  \country{Finland}
}

\renewcommand{\shortauthors}{Epp et al.}

\begin{abstract}
    Social wearables promise to augment and enhance social interactions. However, despite two decades of HCI research on wearables, we are yet to see widespread adoption of social wearables into everyday life. More in-situ investigations into the social dynamics and cultural practices afforded by wearing interactive technology are needed to understand the drivers and barriers to adoption. To this end, we study social wearables in the context of Nordic student culture and the students’ practice of adorning boiler suits. Through a co-creation process, we designed Digi Merkki, a personalised interactive clothing patch. In a two-week elicitation diary study, we captured how 16 students adopted Digi Merkki into their social practices. We found that Digi Merkki afforded a variety of social interaction strategies, including sharing, spamming, and stealing pictures, which supported meaning-making and community-building. Based on our findings, we articulate “Memetic Expression” as a strong concept for designing social wearables.
\end{abstract}




\begin{teaserfigure}
  \includegraphics[width=\textwidth]{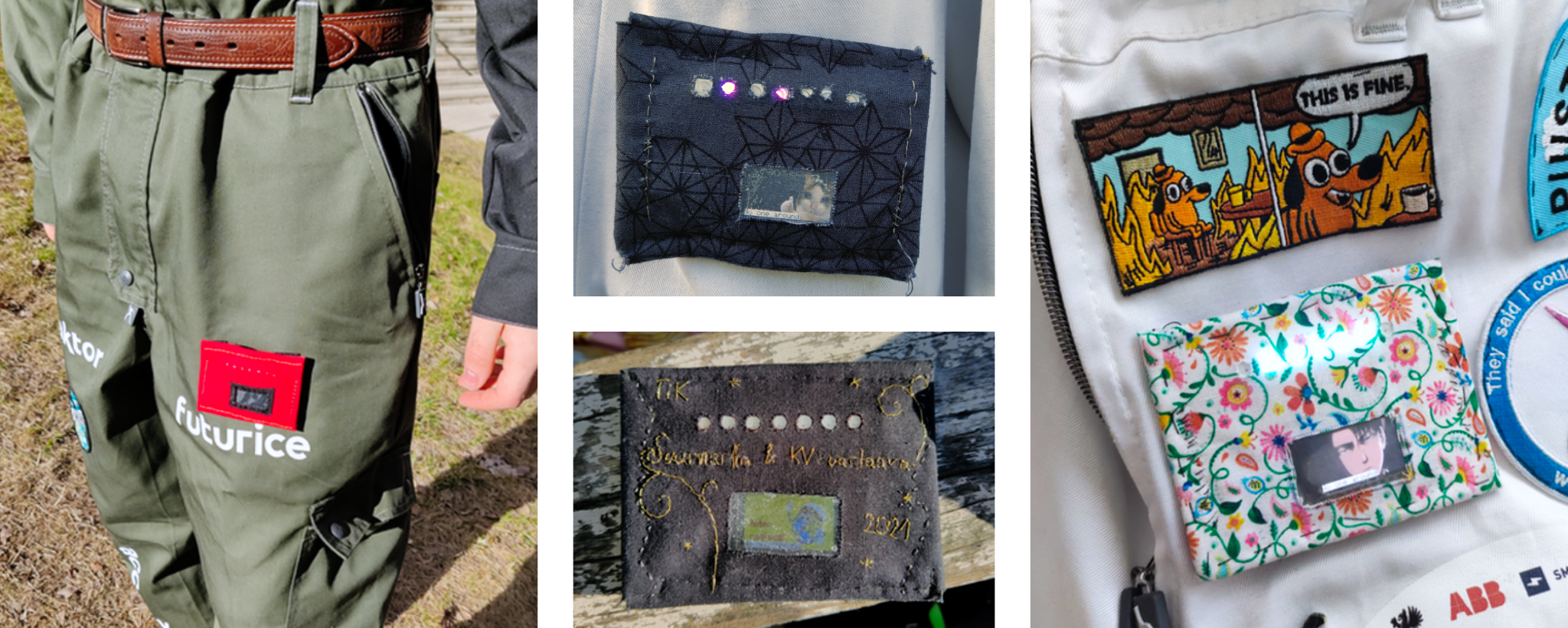}
  \caption{Some of the Digi Merkki that participants created.}
  \Description{A picture divided in four sections by white borders. The left photograph depicts a the pants of an olive boiler suit of a student standing in the sun on a patch of grass. On boiler suits is a red rectangular clothing patch with a small screen with black framing. The top photograph shows a dark grey clothing patch with a some abstract darker line pattern attached to white fabric. A small screen shows the picture of a car raising her thumb and of seven light emitting diodes, two show a pink colour. The bottom photograph depicts a dark grey patch with text and floral patterns embroidered with golden yarn. The small screen shows a picture with an earth cartoonish character wearing a hat on yellow background. The left photograph shows a clothing patches on a white boiler suit. The top solely fabric patch depicts two comic images with a dog sitting in a room on fire with the second picture having a speech bubble that reads “THIS IS FINE”. The lower patch has colourful fabric patterns on whit background and a small screen depicts an anime character. Five bright spots emit white light through the fabric.}
  \label{fig:creations-teaser}
\end{teaserfigure}

\maketitle

\section{Introduction}

Wearables and smart garments have made their way into everyday life. Significant growth of the smart clothes market are expected in the coming years with a myriad of new clothing items that will be connected \cite{researchandmarketsGlobalWearableComputing2020}. Meanwhile, fitness trackers and smartwatches have already become everyday clothing items. Through their immediacy to our bodies, wearables are with us through the patchwork of situations that make up our lives~\cite{mcmillanSituatingWearablesSmartwatch2017}. 

When taking on the form of clothing, wearables become a part of our appearance, and therefore our self-presentation \cite{goffmanPresentationSelfEveryday1959} and their attributes become part of our expressions towards others. Therefore, design scholars have argued that wearables need to be considered in terms of their performative characteristics and consequently included fashion design into the process of creating wearable products \cite{dunneSocialAspectsWearability2014, tomicoNextWaveWearable2017}. As such, wearable design touches the disciplines of HCI, Design and Fashion alike.

Concurrently, HCI scholars have been eager to leverage technology that is co-present in our face-to-face engagements. These co-present technologies promise to enhance our collocated social interactions, augment self-presentation or enable new ways to connect with people~\cite{olssonTechnologiesEnhancingCollocated2019} -- in short, social wearables~\cite{daganDesignFrameworkSocial2019}.
A recent call for a renewed push towards creating such technologies focuses on how wearable technologies can augment social interactions and extend our capabilities \cite{isbisterSuprahumanTechnology2019}. To this end scholars have identified different quality criteria and requirements for design. On the technical level, wearables, both as fashion technologies and as augmentations of social interactions, need the right balance between sensing and actuating \cite{daganDesignFrameworkSocial2019, toussaintKeepingDataGarment2020}.
On the social level, researchers raised questions of power, invasiveness \cite{daganDesignFrameworkSocial2019, toussaintKeepingDataGarment2020}, and social acceptability \cite{daganDesignFrameworkSocial2019, ouversonFashionTechnologyImplications2017}.
However, despite more than two decades of co-present interactive systems, Isbister \cite{isbisterSuprahumanTechnology2019} concludes that the promises of such technologies have not yet successfully transferred into our everyday lives \hl{and commercial wearable still mainly focus on use-cases around health~\cite{jarusriboonchaiCustomisableWearablesExploring2019}}. While it is yet unclear what impedes adoption, it is likely that HCI still lacks a comprehensive understanding of how social technology and wearables constitute people’s everyday social interactions. Most existing research breaks down social interactions into distinct instances of interaction, e.g. in the form of “social ice-breaking”~\cite[e.g.][]{chenOneLEDEnough2016}, without considering the context of people’s lives~\cite{olssonTechnologiesEnhancingCollocated2019}.
Consequently, a better understanding of how to design social wearables that embed into our lives, as well as how they shape social life is needed to identify underlying drivers as well as potentially foster the adoption of wearables for social interactions~\cite{dunneSocialAspectsWearability2014}. 

This gap calls for explorations of technology embedded in people’s everyday life. Here, a perspective on \emph{social practices} is particularly useful, as they are understood as performances of everyday behaviour and building blocks that make up the lives of their enactors~ \cite{shoveDynamicsSocialPractice2012}.
Therefore, we present a research-through-design~\cite{stappersResearchDesign2017} case employing practices-oriented design~\cite{lennekekuijerPracticesorientedDesign2017} that resulted in a novel social wearable embedded into existing practices. The focus of our work was not to design and evaluate a social wearable, but rather to understand people’s practices with social wearables via generative design research~\cite{zamenopoulosCodesignCollaborativeResearch2018}. 

Specifically, our work is situated in the context of Nordic student life, where students practice a rich cultural tradition of wearing colourful boiler suits in private student events and public life (see Figures~\ref{fig:creations-teaser},~\ref{fig:boiler-suit}~\&~\ref{fig:field}). By applying various adornments, such as clothing patches, these suits become an expression of group membership and individuality~\cite{eppIdentitySocialWearables2020}. In this paper, we present the co-design~\cite{zamenopoulosCodesignCollaborativeResearch2018} of a social wearable, Digi Merkki, as well as a \study{field intervention} in the specific cultural context of Nordic university student dress. We investigate how Nordic university students utilise Digi Merkki, an interactive clothing patch, for their socio-cultural practices to gain a better understanding of how a particular community of practice adopts a social wearable, bringing a digital form of expression into a traditionally mostly analogous adornment practice.

From our field study we identified three main themes: %from our qualitative analysis enriched with the log data: 
Digi Merkki shaped our participants’ social practices of adornment. In particular, the openness of our design fostered emergent practices such as dare challenges and spamming. Moreover, participants came up with varying strategies to navigate tensions brought on by the changing practices. Across these themes, we identified internet memes as a helpful tool for participants to mediate meanings when interacting with others through Digi Merkki. 

Based on this, our work makes the following contributions: First, we showcase a \study{design intervention} with a wearable that successfully shaped students’ social practices, at least in the short-term. In particular, we identify memes in digital culture as a mediating concept that fostered integration into existing social practices, as well as facilitating emerging novel practices. 
%We identified memes in digital culture as a prevalent concept structuring the practices around a social wearable in Finnish student culture
Second, we articulate Memetic Expression as a novel strong concept for the design of social technology to foster identity and community. Taken together these contributions may inform researchers and designers in understanding and guiding the adoption of social wearables.

\section{Related Work}

\todo{filler sentence}

\subsection{Augmenting Face-to-Face Social Interactions}

Recently, HCI scholars have argued for renewed efforts in using interactive technologies for augmenting collocated social interactions \cite{isbisterSuprahumanTechnology2019, olssonTechnologiesEnhancingCollocated2019}. This call follows an over two-decades-long history of researching technology that suits interpersonal interactions in the “same place, same time”. Early works studied wearables \cite{borovoyMemeTagsCommunity1998, falkBubbleBadgeWearablePublic1999, paulosFamiliarStrangerAnxiety2004} and public displays \cite{mccarthyAugmentingSocialSpace2004} to present personal information to collocated people. With the advent of mobile devices, researchers investigated self-expression~\cite{perssonDigiDressFieldTrial2005} and picture sharing~\cite{luceroMobiComicsCollaborativeUse2012} by leveraging the proximity of people for engagement through technology. In recent years, research efforts have concentrated on understanding uses of personal displays for public viewing \cite{colleyExploringPublicWearable2020, jarusriboonchaiIncreasingCollocatedPeoples2016, kleinmanExploringMobileDevices2015, kytoAugmentingMultiPartyFacetoFace2017}.
Especially, advances in fabrication and smart textiles have led to the exploration of wearable displays and expressive garments to attract attention~\cite{chenOneLEDEnough2016, dierkAlterWearBatteryFreeWearable2018, harjuniemiIdleStripesShirtWearable2020, pearsonItTimeSmartwatches2015}, display personal information~\cite{ashfordExplorationResponsiveEmotive2019, colleyExploringPublicWearable2020, colleySmartHandbagWearable2016, harjuniemiIdleStripesShirtWearable2020, hartmanMonarchSelfExpressionWearable2015, pearsonItTimeSmartwatches2015}, or other social signalling \cite{Devendorf2016, hartmanMonarchSelfExpressionWearable2015, mackeyCanWearThis2017}.
Another strand of research has looked at wearables as more proactive tools of changing collocated social interactions, like secret nudging~\cite{katehartmannNudgeables} or gaming \cite{isbisterInterdependentWearablesPlay2017, burukDesignFrameworkPlayful2019}. Both strands can be considered under the term  “social~wearables”~\cite{daganDesignFrameworkSocial2019}.

Despite these efforts and advancements in wearable computing, widespread adoption of these social wearables has remained limited, despite computer-mediated communication permeating everyday life. This disparity was attested by Olsson and colleagues in their literature review of technologies for collocated social interaction \cite{olssonTechnologiesEnhancingCollocated2019} and critiqued in Isbister’s suggestion for “supra-human technology” \cite{isbisterSuprahumanTechnology2019}. 
Existing works have used approaches centred around a design artefact~\cite{colleyExploringPublicWearable2020, Devendorf2016, dierkAlterWearBatteryFreeWearable2018, katehartmannNudgeables, colleySmartHandbagWearable2016, harjuniemiIdleStripesShirtWearable2020},  data~\cite{chenOneLEDEnough2016, colleyExploringPublicWearable2020,  jarusriboonchaiIncreasingCollocatedPeoples2016},  or the users’ proximity~\cite{chenOneLEDEnough2016,  jarusriboonchaiIncreasingCollocatedPeoples2016}. Only few studies~\cite{daganDesigningTrueColors2019, mackeyCanWearThis2017} have investigated the social settings, practices and rules in place. This deficit corresponds with Olsson and co-authors’ observations for all collocated social interactions technology~\cite{olssonTechnologiesEnhancingCollocated2019}.
This research gap is particularly evident when we consider that the social relationship between people is the strongest factor for interpersonal engagement \cite{wieseAreYouClose2011} and social influence a dominant factor for the adoption of collaborative technology~\cite{olschewskiCollaborationTechnologyAdoption2018}. Hence, a deeper dive into everyday settings and social practices is necessary to gain insights into the drivers underlying the adoption of wearable social technology.

\subsection{Designing Social Wearables} \label{sec:related-work-designing}

Wearables can indeed be adopted in our everyday lives while integrating with our social interactions, as exemplified by findings on smartwatch use~\cite{mcmillanSituatingWearablesSmartwatch2017}. However, in the previous section, we identified a gap of technology that actively augments social interactions. Dunne and colleagues \cite{dunneSocialAspectsWearability2014} have already identified aesthetics, social identity and cultural norms as factors of social acceptability of wearables. 
As these norms and identities change over time for every individual, Dunne and colleagues conclude that a checklist for achieving the social acceptability of wearables is not feasible.
Instead, they suggest several guiding questions. For example, “What mechanisms are available to mediate that acceptance process?”~\cite{dunneSocialAspectsWearability2014} 
Similarly, privacy is an often quoted factor for the successful adoption of social wearables~\cite{daganDesignFrameworkSocial2019}. 
However, Dourish~\&~Bell~\cite{dourishRethinkingPrivacy2011} suggest privacy as a dynamic concept of negotiation for ubiquitous technology to “support the human social and cultural practices through which [privacy and related concepts] are managed and sustained.”
Over recent years, other works have contributed guiding principles. 
In their work on textile screens, Devendorf and colleagues~\cite{Devendorf2016} propose ambiguity as a tool to allow for open experiences. 
This aligns with Fallman’s~\cite{fallmanNewGoodExploring2011} assertion that for experiences that should be “lasting or even grow …, usability may be counterproductive”. 
Even more so, interactions that require effort can create meaningfulness and therefore be a source for design, especially in socially engaging systems~\cite{kellyDemandingDesignSupporting2017}.

Similarly, work on social wearables proposed openness in the functioning of systems. Dagan and colleagues \cite{daganDesignFrameworkSocial2019} suggest that the interplay of actuation and sensing might be reconfigured by the wearer, as a promising approach. Further, social wearable scholars made use of “strong concepts”~ \cite{hookStrongConceptsIntermediatelevel2012} to formulate guiding principles. With “interdependent wearables”~\cite{isbisterInterdependentWearablesPlay2017} and “synergetic technologies”~\cite{daganSynergisticSocialTechnology2021}, Isbister and colleagues articulated strong concepts that can be leveraged for social wearables and beyond.

Outside of the field of HCI, the adoption of wearables is also studied in marketing research. Kalantari~\cite{kalantariConsumersAdoptionWearable2017} suggests user-centred methods and that users “actually wear the devices” when inquired about adoption. Ideally, such studies should investigate how users “develop attitudes towards wearable technologies over time.” This aligns with Olsson et al.’s~\cite{olssonTechnologiesEnhancingCollocated2019}  proposal to focus on distinct socio-cultural settings. 
Indeed early works in the field of “same place, same time” technologies proposed design for culture and community~\cite{borovoyFolkComputingRevisiting2001, kortuemWearableCommunitiesAugmenting2003}. 
However, HCI scholars have rarely picked up on this avenue. 

\subsection{Adornment, Fashion and Digital Expression} %\subsection{Smart Garments, Fashion and Digital Expression}

In recent years, fashion design has looked into wearables in the form of “fashionable wearables”~\cite{gencRecommendationsDesigningFashionable2018, tomicoNextWaveWearable2017} and “digital expression” in clothing~\cite{mackeyCanWearThis2017}. 
Here, researchers try to understand how digital garments might integrate their technical functions into the social functions of dress. 

Adornment long has been understood as a tool for an individual to demarcate personality~\cite{simmelSociologyGeorgSimmel1950}. 
In her seminal book “Adorned in Dreams” Wilson~\cite[][p.3]{wilsonAdornedDreamsFashion2003} unites adornment, dress and fashion as human activities that do not just express superficiality but actually “play symbolic, communicative and aesthetic roles.” 
From the current understanding in fashion studies, we can see wearables as part of the social system of reproducing symbols and identity through consumption and display~\cite{entwistleFashionedBodyFashion2015, barnardFashionTheory2014}. 
Work in fashion design on wearable technology~\cite{gencRecommendationsDesigningFashionable2018, tomicoNextWaveWearable2017, mackeyCanWearThis2017} suggests exploring people’s lived experience in “genuine social contexts” to identify “emerging patterns”~\cite{mackeyCanWearThis2017}. 
In “same place, same time” literature, few works have evaluated their design for effects on the social practices of participants~\cite{olssonTechnologiesEnhancingCollocated2019}. 
Hence, clothing practices and adornment are particularly promising social practices to explore wearables that are meant to augment our social interactions. 

\begin{figure}[htb]
  \centering
   \includegraphics[width=0.95\linewidth]{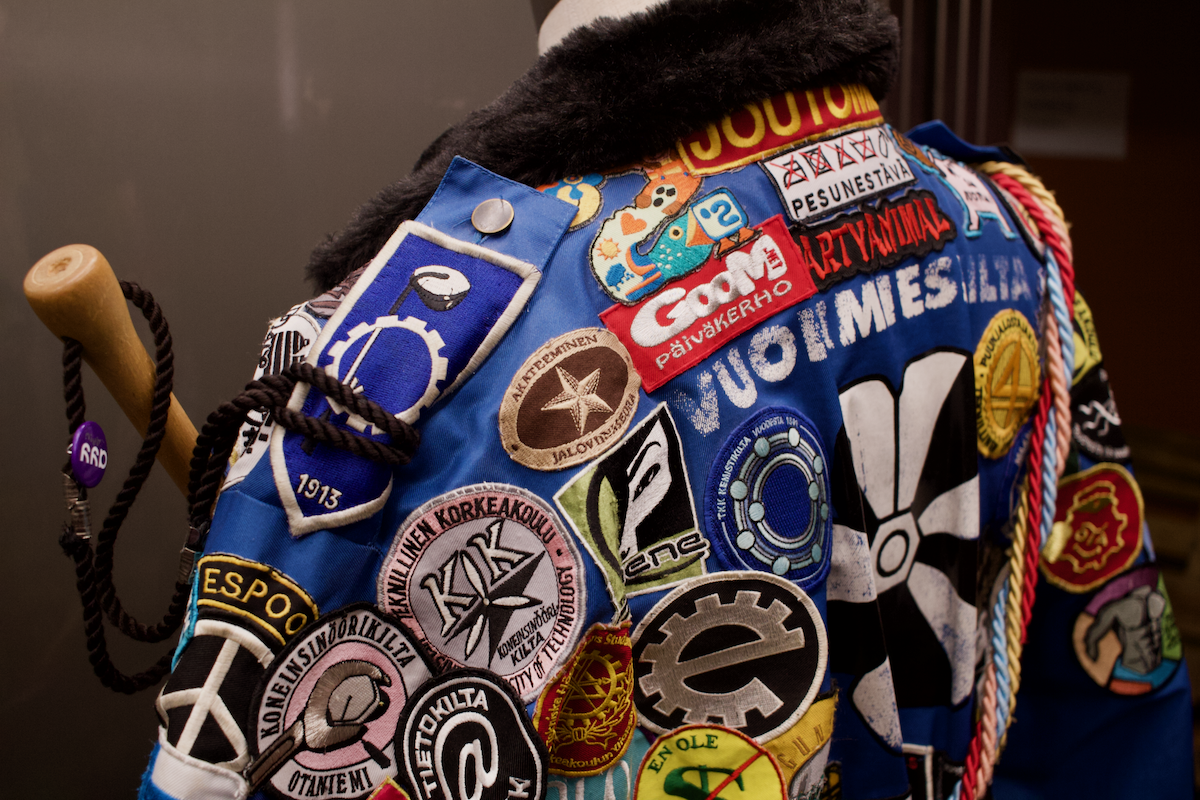}
  \caption{The boiler suit of an active student after finishing their studies.}\label{fig:boiler-suit}
  \Description{A photograph showing the upper back part of a boiler suit. The blue suit is almost completely covered in textile patches in different colours and shapes with Finnish writings of different associations and logos. Only the centre back area showing the name and logo of the subject associations. Additionally the suit has a fur collar and decorative ropes of various colours hanging form the right shoulder.}
\end{figure} 

\section{Nordic Student Culture and the Practice of Adorning Boiler Suits} \label{sec:case}

Fashion is a tool used by people to form distinct social groups~\cite{wilsonAdornedDreamsFashion2003}. University students often follow certain styles, even as a form of tradition, e.g. letterman jackets in North America \cite{wikpediaLettermanSports2021}. In European countries fraternities have followed particular cultural dress codes for centuries. Nordic students, for instance, developed a tradition of wearing colourful boiler suits~\cite{wikipediaStudentBoilersuit2021}. Student associations in Finland, Sweden and Canada provide their members with these suits as a form of affiliation~\cite{wikipediaStudentBoilersuit2021}. In Finland, students dress in their boiler suits for different occasions: parties and ceremonies, especially around spring carnival (fi: \textit{Vappu}), but also educational events and excursions. These colloquially named “student overalls” (fi: \textit{opiskelijahaalarit}) have a particular colour for each student association and are customised by each student individually.

We based our work on this particular socio-cultural context, as it serves as a clear delineation of social practice. Practices depend on the interconnections between different elements, which Shove~et~al.~\cite{shoveDynamicsSocialPractice2012} categorise as materials, competences and meanings. As connections form and break between these defining elements, practices emerge, persist and disappear. Next, we describe the practice of adorning boiler suits in more detail based on previous ethnographic accounts~\cite{eppIdentitySocialWearables2020}, as well as our own observations and experiences. %into their constituting elements.

\subsubsection{Materials}

\hl{The dress oriented practices include several clothing items as materials:}
The oldest clothing item connected to this practice is the student cap. \hl{Nordic students traditionally wear} white student caps during the summer period starting with the Vappu carnival celebrations on the eve of May the first.
Technical students (colloquial Finnish: “Teekkari”) earn the right to wear a specific Teekkari cap.
These technical students also first introduced boiler suits as a student outfit.
Nowadays most subject associations hand out overalls or lab coats as uniforms for their members, each with a specific association colour.
Central to the adornment practice are clothing patches.
These patches are handed out by associations, in events, as promotional materials by companies or sold by patch dealers.
Students collect a large amount of patches, so it is quite common that space on the boiler suit runs out (see Figure~\ref{fig:boiler-suit}). Therefore students have a pocket full of loose patches to sew on later or trade.
Several other adornments are also frequently used, like drinking cups, belts, ropes, pins or custom-made collars and hoods.
\hl{Especially for the technological students these materials include electronics reflecting their field of study.}

\subsubsection{Competences}

In the first week of studies new students are introduced to the traditions of the student community and subject associations.
The associations even give dedicated lectures on the particular clothing traditions, including the boiler suits. 
Due to mandatory military service most men already have knowledge of military traditions, which help them to understand ranks, structures and heraldry symbols inside these student associations. 
Most Finnish students already know how to do basic sewing due to housekeeping education in primary school. 
Students gather other competences supporting their adornment practices from subjects or topical associations.

\subsubsection{Meanings}

Most meanings associated with this practice revolve around the identities of the students. 
By adorning their suits, they generate a personal presentable history, which represents their status and membership. 
Consequently a lot of themes in this community deal with community and brotherhood.
Being a student means being part of youth culture, as one lives the first time by themselves. In so far, testing out limits, e.g. alcohol, intimate relationships, is tightly intertwined with the adornment practice itself, for example, legs of the boiler suit may be exchanged with intimate partners and arms with friends.
While this community cherishes traditions, they also celebrate reinventing them. In general self-made adornments are valued highly, exemplified by the rule to sew on the patches, never glue. Consequently local memes and their distribution can constantly be seen with ever new patches referencing existing memes and student life (see the adaptation of the meme “This is fine”\footnote{This Is Fine. Know Your Meme. Retrieved January 10, 2022 from \url{https://knowyourmeme.com/memes/this-is-fine}
} in the right picture in Figure~\ref{fig:creations-teaser}).

\section{Co-Designing an Interactive Clothing Patch}

In the following section, we describe our collaborative design process that led to our \study{field intervention}. We illustrate our design activities and the characteristics of the research prototype. %that was then used to study the case of digital adornments.

\begin{figure}[htb]
  \centering
   \includegraphics[width=0.95\linewidth]{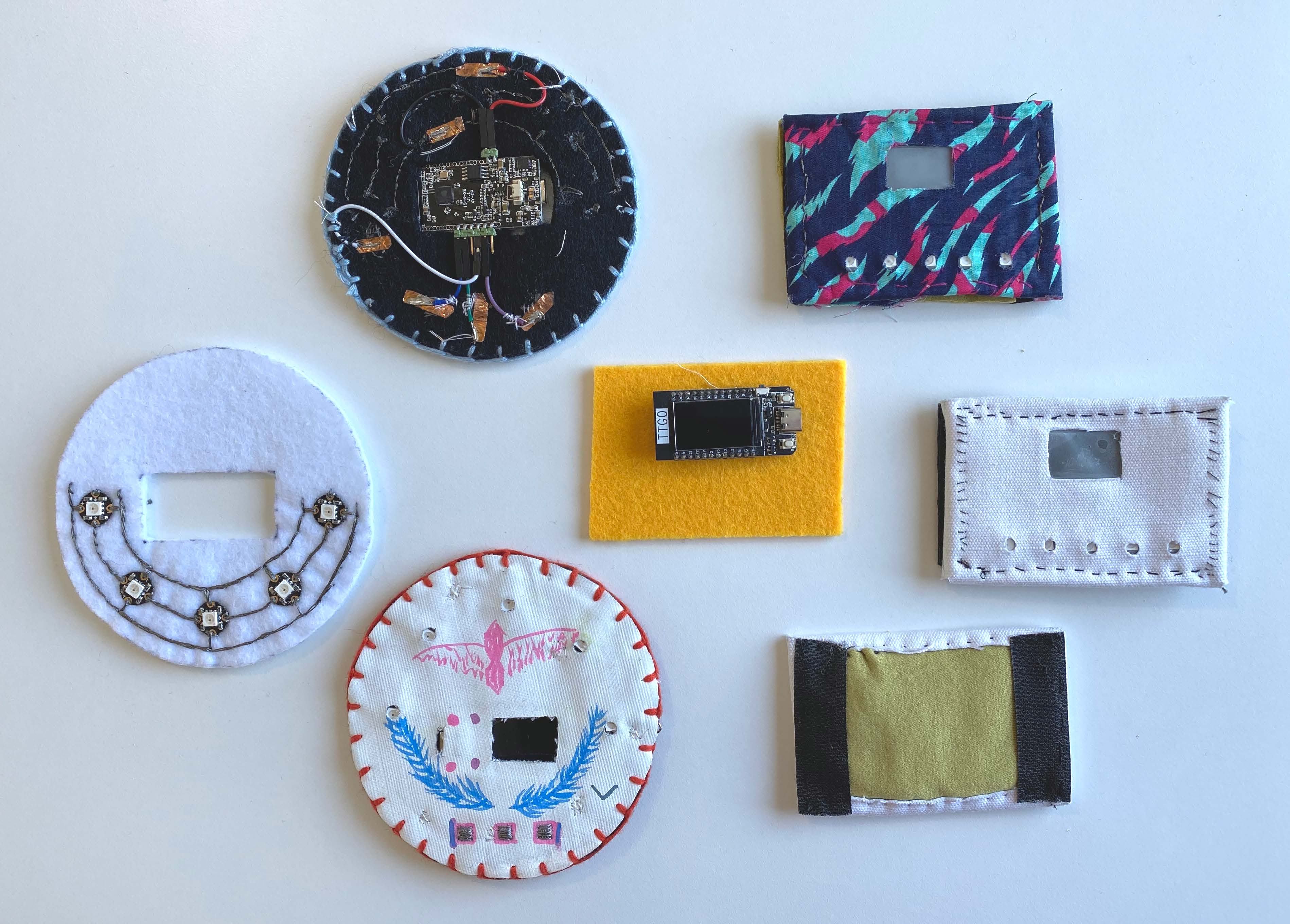}
  \caption{Different stages of the prototypes during our co-design activities.}\label{fig:prototypes}
  \Description{The photograph depicts seven different patches. Three objects are circular with a black one exposing some cabling and a rectangular micro-controller board, a white one exposing some electronic components connected through embroidered stitches and a rectangular hole, and another white one with some drawings, a black cutout and three embroidered spots at the bottom. In the centre of the photograph is a rectangular piece of yellow felt with a micro-controller placed atop that has a small screen. The rectangular patches on the right show two patches with a window for a small screen and a line of holes, one in white and the other a blue, cyan and pink animal print. The bottom patch is plain olive green with black stripes one the left and right.}
\end{figure} 

\subsection{Co-design Process}

Our motivation was to understand the adoption of social wearables situated in people’s practices. This practices-oriented approach~\cite{lennekekuijerPracticesorientedDesign2017} deals with a speculative future.
We choose the generative, collaborative research practice of co-creative design as it fosters “learning from the collective creativity” of the community to innovate~\cite{zamenopoulosCodesignCollaborativeResearch2018}.
This co-design approach is generative in developing and studying prototypes~\cite{sandersProbesToolkitsPrototypes2014}.
Therefore, the design process needed to result in a wearable research prototype for augmenting social interaction that embeds in the local cultural practices.  
Here, we chose to design for possibility~\cite{desmetHappinessPossibilityDrivenDesign2012} not to solve a problem, and built on the community’s existing assets, instead of its needs~\cite{wong-villacresReflectionsAssetsBasedDesign2021}.
This position shifted the perspective towards a potential future this community desires.
In our process we facilitated co-design tasks, implemented various prototypes, and iterated based on participant feedback and testing. 

The design group consisted of the first author, who was responsible for facilitation and prototype implementation and up to four more HCI researchers and designers from the same institution with varying involvement. Additionally, we invited two students with extensive knowledge of the local student culture as co-designers into our design group. Both co-designers had extensive proficiency in crafting student outfits, in organising community events in the local student culture, and student traditions, and one of the co-designers was a cosplayer~\cite{lamerichsCostumingSubcultureMultiple2014}, i.e., they create and wear their own costumes representing popular media characters. Together with the first author these two students were participating in the design process throughout. %In the following section “we” refers to the design group

We adapted the design framework for Social Wearables by Dagan and coauthors \cite{daganDesignFrameworkSocial2019} to guide the design decisions. The framework encompasses five aspects of social wearables for design: Sensing, Actuating, Sensing-Actuating Interplay, Personal and Social Requirements, and Social Acceptability. We added two categories that did not focus on interaction but material characteristics. The factor “form” asked for material characteristics as well as clothing style and aesthetics. The factor “environment” asked for endurance requirements and context of use, e.g. “water resistance”.\footnote{See the template in the supplementary material.} We derived these additional requirements from existing frameworks of wearability \cite{dunneSocialAspectsWearability2014, mottiHumanFactorsConsiderations2014}.

%As highlighted in section~\ref{sec:case} the adornment practices revolve around developing identity.
%We formulated the following design brief: “Build interactive and expressive overalls which you can use with others this [spring carnival] and beyond”.

We formulated first ideas for Digi Merkki during a co-design workshop on social wearables for eight students facilitated by two designers.
This workshop was for design and engineering students to work on individual ideas for wearables based on our revised social wearables design framework. 
The workshop was informed by design concepts for Nordic students identified in our ealier work~\cite{eppIdentitySocialWearables2020} as inspirations. 
% had conducted speculative co-design to generate design concepts for wearables for Nordic Students. 
Our design group saw potential in the concepts “Personal Displays For Ice-breaking”, “Displaying Group Belonging Dynamically”, “Interactive Badges for Collecting”, “Grabbing Attention”, and “Status Display” (see ~\cite{eppIdentitySocialWearables2020} for more details).
Therefore our design process aimed for a social wearable that brings people together and supports them in self-expression.

Based on the community practices revolving around developing identity, the concepts and the feedback from students during the workshop, we decided on the main functionality and piece of clothing of our social wearables in our group of co-designers.
The core functionality of Digi Merkki aimed for “enabling handshakes”, “displaying social ties”, and “self-expression” as a merge of the concepts above. This initial idea aimed at directly integrating the technology into the overalls themselves.

We changed the design to non-physical handshakes to offer interactions with strangers, which also helped in the context of changing physical distance regulations due to the COVID-19 pandemic. We then decided on the form of an interactive clothing patch that connects to other patches nearby, which enabled interaction over bigger distances and more flexibility in how students would use the wearable on their overalls.
Together with our co-designers, we identified requirements based on these goals: Digi Merkki needs to integrate into the overall and not become an additional device; visualisations must be visible beyond 1.5 meter distance; the patch needs to be water-resistant for outside use. 

After producing the first rapid prototype (circular patch in Figure~\ref{fig:prototypes}), we probed three devices in a \study{week-long field test} with three people (two researchers from the design group, one external). This test made us explore the interaction concepts, the wearability and technical implementation.
Based on the findings, we refined the interaction design and aesthetics in another co-design session.
Initially, users performed the non-physical handshake by entering a five-digit cypher. We realised that trading pictures simulates a handshake and switched to a time-based interaction.
The first prototypes had touch surfaces that resembled push buttons. However, those interactions made the device appear as a mobile device instead of a wearable embedded into clothing. To enable more embodied interaction, capacitive sensing was embedded into the seams, so users can use their whole hands to cover parts of the patch.
We also changed the dimensions to the final format (rectangular prototypes in Figure~\ref{fig:prototypes}) and decided on implementing a semi-flexible attachment with hook and loop fasteners to allow a variety of body positions. This way we managed to create something integrated enough to be considered part of the overall, while still allowing for easy access to the electronics.

We then tested the final functionality in a four-person event (one researcher from the group, three externals). This mainly contributed to stability improvements. After that, the co-design focused on structuring the \study{intervention} around activities for the students, not just the prototype itself as we moved on to testing the wearable with students in-situ. We revisited the production process (see Figure~\ref{fig:prototypes}) to make it easier to multiply and copy Digi Merkki. We included options to customise the physical (front layer fabric) and digital (personal pictures and LED colour) aspects of Digi Merkki to the students preferences. The final product was a creation kit with the electronics ready-to-use and instructions for producing a personal patch (see Figure~\ref{fig:kit}). \hl{Here, we chose involving handcrafts for enabling personalisation and leveraging pre-existing skills~\cite{perner-wilsonHandcraftingTextileInterfaces2011}.}

\begin{figure}[htb]
  \centering
  \includegraphics[width=\linewidth]{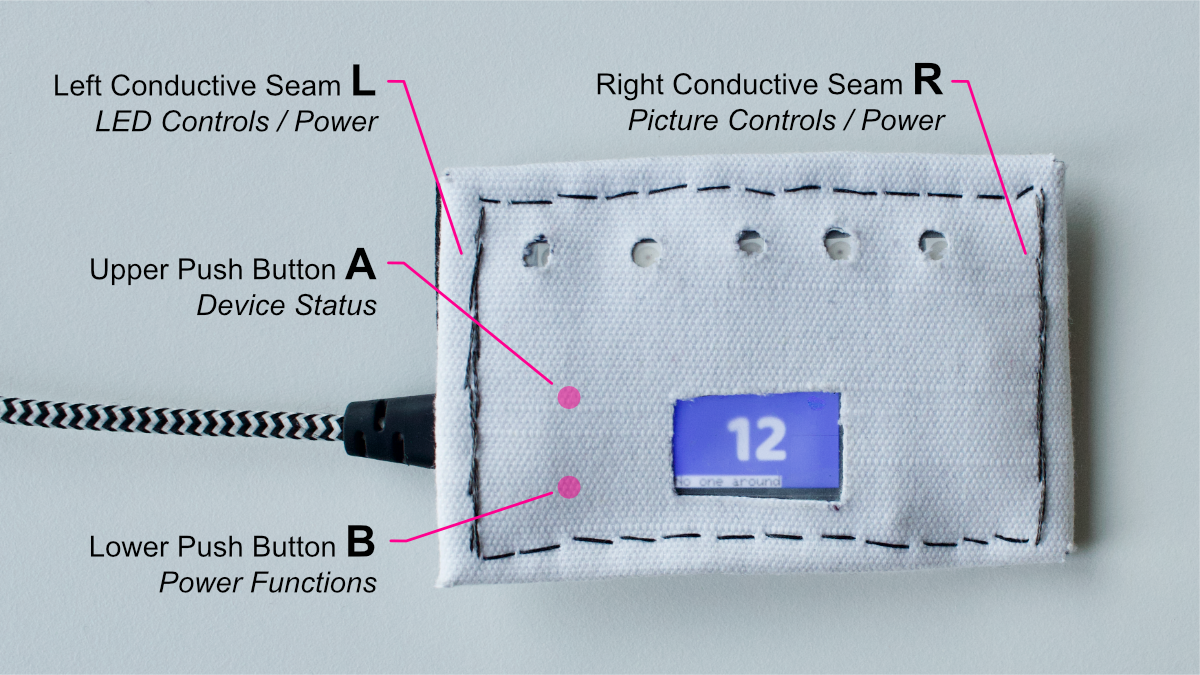}
  \caption{A Digi Merkki prototype showing the final functionality while charging.}\label{fig:controls}
  \Description{A photograph showing a white rectangular clothing patch on grey background. A cable comes from the left underneath the patch. The patch has grey seems along all four sides. On the top is a line of five holes. On the bottom is a rectangular hole exposing small screen showing a written twelve in white on purple background. Four pink arrows point towards the patch with black text written to each end. One points to the left seams and reads “L: Left Conductive Seam, LED Controls / Power”, one to the right seems and reads “R Right Conductive Seams, Picture Controls / Power”, two point to pink overlay dots halfway between the small screen and the outer left edge, with one at the height of the upper edge of the screen reading “A Upper Push Button, Device Status” and the other one at the height of the lower edge of the screen reading “B Lower Push Button, Power Functions”.}
\end{figure}

\begin{figure*}[tb]
  \centering
  \includegraphics[width=0.8\textwidth]{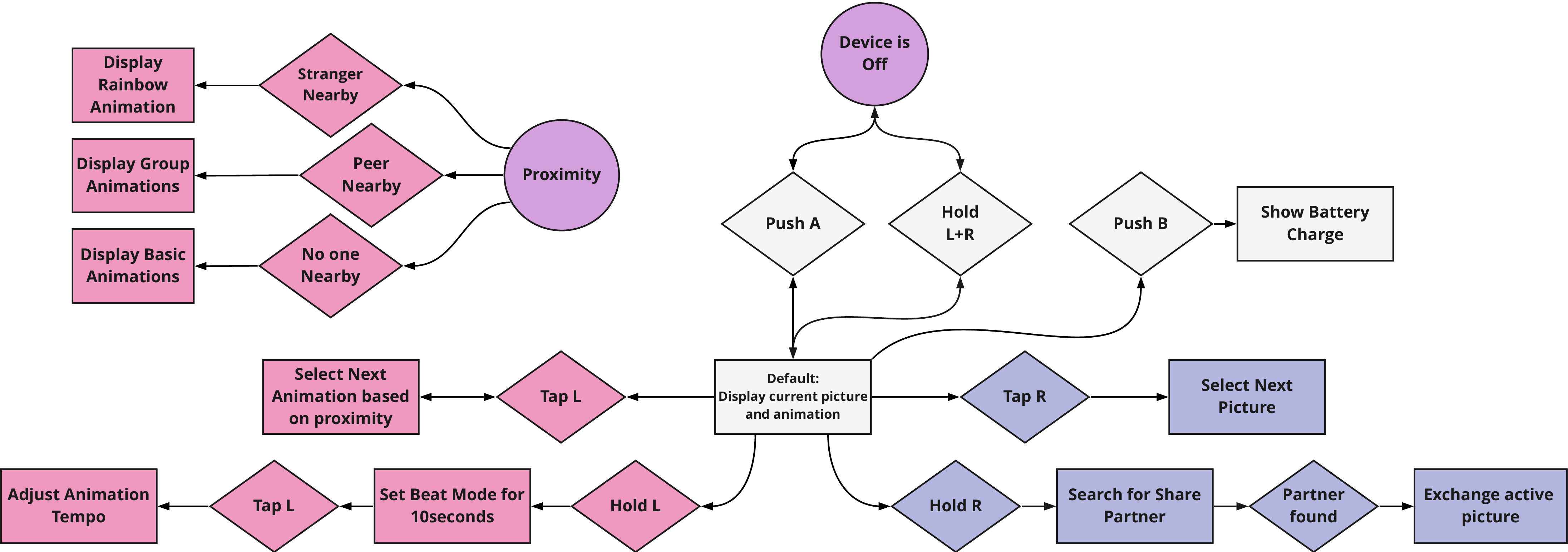}
  \caption{Flow diagram of Digi Merkki depicting the operation of the light animations in red and the picture trading in blue.}\label{fig:flow}
  \Description{A flow chart of the user inputs and states of the interactive patch with different section of the chart coloured with different shades based on main functionalities.
  From the topmost item "Device is Off" in pale purple, both, "Push A" and "Hold L+R" lead down to the central element "Default: Display current picture and animation".
  From there "Push B" leads off to "Show Battery Charge"in the top right corner. All these elements are in grey.
  To the right side of the chart in greyish purple lead the states of the picture functions away from the default state. "Tap R" leads to "Select Next Picture" and "Hold R" leads to "Search for Share Partner" then "Partner found" following "Exchange active picture".
  To the left side of the chart in pale pink lead the states of the LED animation functions away from the default state. "Tap L" leads to "Select Next Animation based on proximity". "Hold L" leads to "Set Beat Mode for 10seconds" then to "Tap L" and finally "Adjust Animation Tempo".
  Detached from the rest of the chart in the top left corner is the pale purple starting point "Proximity". From there three branches in pale pink go off. "Stranger Nearby" leads to "Display Rainbow Animation", "Peer Nearby" leads to "Display Group Animations", and "No one Nearby" leads to "Display Basic Animations".}
\end{figure*} 

\subsection{Scenario and System Design}

%[diagram with an overview of characteristics]
%[Add attributes from Dagan et al. framework]
%[A Design Framework for Playful Wearables. (Buruk et al., 2019)]

Looking at Digi Merkki as a mobile experience for collocated interactions, we can describe it from a temporal, spatial, social and technological perspective~\cite{lundgrenDesigningMobileExperiences2015}. Without clear goals or a time frame Digi Merkki allows for individual pace. Patches connect to each other in an approximately 50-meter radius. The proximity to other users enables social interactions through the devices. The primary features are synchronous visualisations based on proximity and social ties and trading digital images with collocated users. These technological functions rely on the coordination of activities between participants. Apart from the relation towards other users and their decisions, there are no external triggers from the environment.

%\subsection{Sensing, Actuation and Sensing-Actuation-Interplay}
The final patch consists of a screen, LED lights, WiFi for connectivity, and capacitive seams and push buttons to control the device with a size of approximately $65 x 95$ mm and a thickness of about $11$ mm.
The patch will alert the user of nearby devices by changing to a specific LED animation (see Figure~\ref{fig:flow}): Each participant’s lights have a user-defined colour. Being around people who are part of your friend group activates more complex albeit single-colour animation patterns. If strangers are close by, the patch lights up in a rainbow pattern.

The patch senses user input through capacitive touch embedded into the seams. As shown in Figure~\ref{fig:controls}, there are two touch-sensitive seams to use the patch. The right seam (R) controls the pictures on the screen, the left seam (L) controls the LED animations and touching both at the same time controls power. Additionally, two push buttons (A, B) are under the fabric next to the screen for secondary functions and fallback power control.
A single tap on L or R selects the next animation or picture, respectively. Additionally, users can change the speed of all animations by holding and then tapping L. Adjusting the speed also suynchronises the speed on the patches of close-by users.
To trade pictures, two users have to hold down R for five seconds simultaneously. This will exchange the pictures currently selected in their respective screens. 

\subsection{Technical Implementation}

All devices follow the same technical design. This allowed us to reproduce devices in sufficient numbers to function collectively. As a basis for the computational requirements, we choose development boards based on the ESP32 chip to handle logic and establish connectivity. We decided on WiFi and a mesh network topology to allow connections between different aggregations of collocated users and the use of a high-level protocol to support our rapid prototyping. For user input, we used capacitive sensing in the seams of the device. After testing different setups and textile force sensors, we chose those to enable robust low fidelity interactions. For output, the micro-controller has an embedded $240 x 135$ pixel TFT screen. Additionally, we attached a flexible strip of addressable colour light-emitting diodes. The device is powered by a 1000mAh Lithium-Ion battery attached to an internal charging controller on the micro-controller.

\section{Field Intervention}

The %research-through-design study’s 
main goal of this work was to explore potential drivers of adoption of social wearables from an in-situ perspective on practices. Therefore, the focus of our work was not to design and evaluate a social wearable but rather to understand changing user practices.
We here understand adoption as the Nordic students’ utilisation of a social wearable in their cultural practices. To this end, we conducted a \study{field study} adjunct to our design process. This \study{field study} focused on exploring the effects on students’ practices of creating and using an interactive clothing patch.

We produced 20 co-creation kits with the same functionality and distributed them to a mix of individuals and groups. First, participants customised and finished their devices in remote group workshops or alone. Then we captured their use and interactions with an elicitation diary study over two weeks with four organised gatherings. Finally, we interviewed each participant indivually for their experience. The university’s ethics committee assessed and approved the study.

\begin{figure*}[t]
  \centering
  \includegraphics[width=\textwidth]{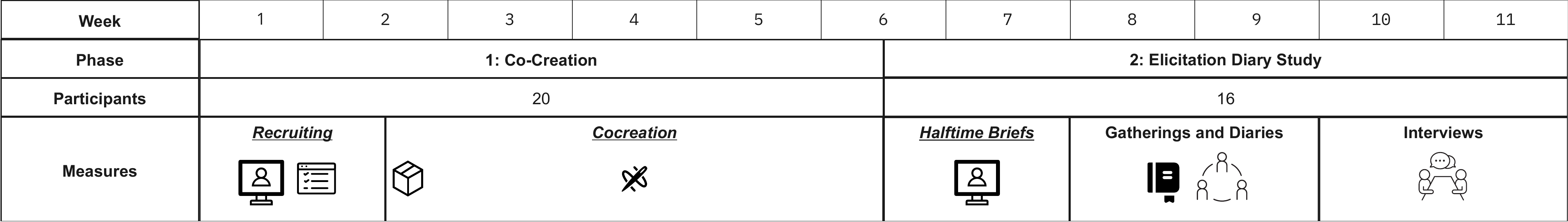}
  \caption{Timeline of the \study{field study} over eleven weeks. The activities (written in italics and underlined) were held remotely until the seventh week.}\label{fig:timeline}
  \Description{Timeline of co-creation from week 1 to week 6 and elicitation diary study from week 6 to week 11.}
\end{figure*} 

\subsection{Research Questions}

As our unit of analysis was students’ practices~\cite{kuuttiTurnPracticeHCI2014} rather than individual momentary interactions with Digi Merkki, %we focused on how students made use of the intervention. By introducing a new element in Digi Merkki as a new material, we expected practices to adapt. 
% As our unit of analysis were practices we wanted to shape our perspective beyond the interactions with device.
%Hence, 
we formulated the following research questions to guide data collection and analysis:

\begin{itemize}
\item RQ1: How do individual students utilise Digi Merkki with different strategies?
\item RQ2: What measures do participants perform to adapt their self-expression practices and social interactions to the constraints introduced by Digi Merkki? 
\end{itemize}

%The design of the creation kits and Digi Merkki aimed for individual appropriation. Therefore, 
RQ1 helped us to keep an individual perspective. 
RQ2 looks at the relationship between the changes in practices and the specific characteristics of Digi Merkki. This perspective should help identify insights on the barriers and limitations of adoption in people’s practices.  
%RQ2: We understood constraints, not as limitations, but as structuring properties of Digi Merkki as a material in the students practices. In both cases we were more interested in utilisation then artefact use.

\subsection{Participants}

For our study, we recruited 20 students (13 men, five women, one non-binary person, one person chose not to disclose their gender) between 19 and 26 years (average 22). To have a mixed sample of the local student population, we recruited students from different disciplines. The participants spread across different study years (seven first-year, seven second- to third-year and six fourth year and later) and fields, from computer sciences (12), other STEM fields (5), as well as pedagogy (2) and design (1). All students were active in student associations like subject guilds or interest clubs (e.g. crafts or video games). They attended multiple association events a year, with nine participants holding active chairing roles in their respective associations. We recruited individuals and groups of friends to have a mix of different social ties between our participants.
%[Backgrounds in wearables, design and student culture]
Each participant received ownership of their \hl{creation} kit and their version of Digi Merkki. Four participants did not join the elicitation diary study due to time constraints. For the additional time spent on the data collection, each participant of the remaining 16 participants received two cinema vouchers with an approximate value of 15 USD.

\subsection{Procedure}

The primary method to collect data on our research questions was a \study{field study} using a \study{design intervention}.
As shown in Figure~\ref{fig:timeline}, the \study{field study} consisted of two main phases the co-creation and the elicitation diary study:

\begin{enumerate}
\item Participants remotely created personal interactive clothing patches based on our electronic textiles co-creation kits.
\item Participants wore Digi Merkki and reported their experience in a two-week elicitation diary and subsequent individual interviews. 
\end{enumerate}

\subsubsection{Recruitment and Questionnaire}

We advertised the study in local student social media channels. After registering their interest, we invited students to a personal video call. The personal calls helped brief participants about the procedure and establish a personal connection between researchers and participants. This connection was essential for the personal character of the interviews. Participants also received all necessary study documents electronically before the study. 

After this briefing, students could sign up by filling in a questionnaire. The questionnaire captured standard demographic information and more detailed factors of the student’s participation in the local cultural activities. Finally, participants provided three digital images and a colour-code for their personalised interactive clothing patch.

\begin{figure}[htb]
  \centering
  \includegraphics[width=\linewidth]{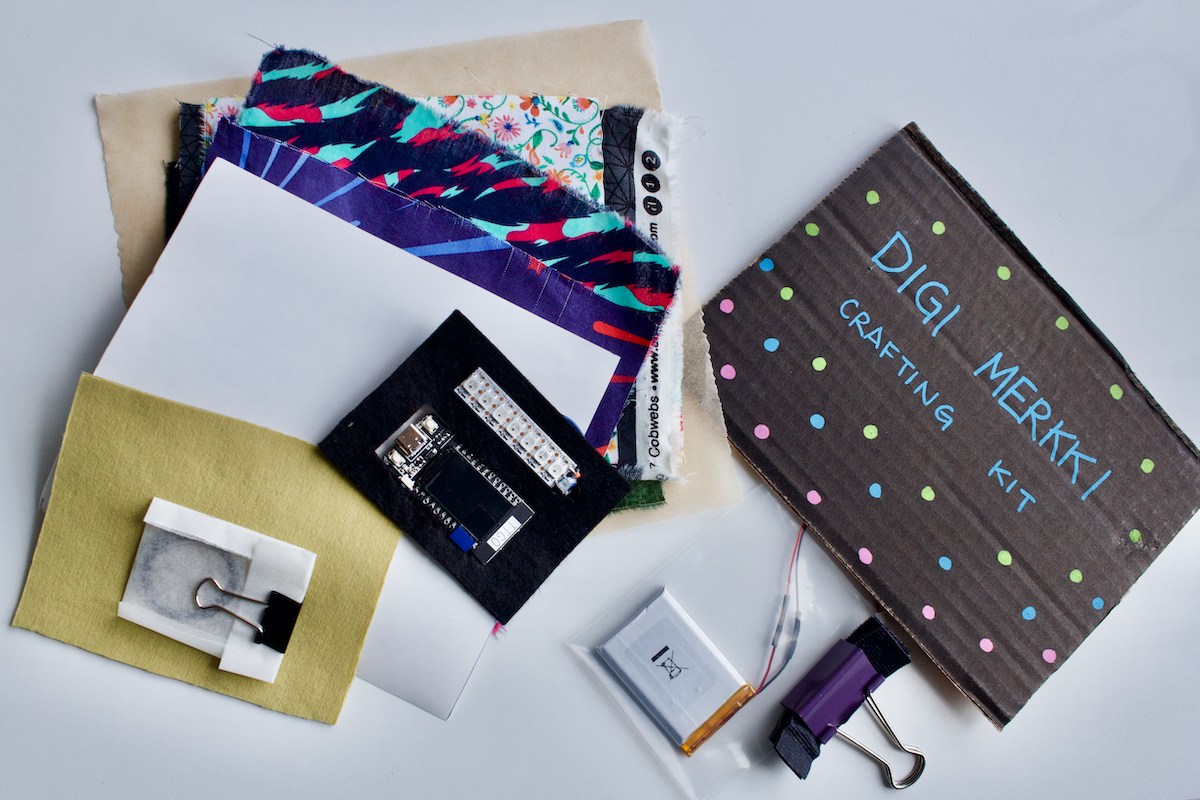}
  \caption{Contents of the co-creation kit that each participant received.}\label{fig:kit}
  \Description{The picture is showing the contents of the Digi Merkki creation kit layed out on a grey surface. The kit contains the electronics patch prototype (a rectanuglar piece of yellow felt with a microntroller board and a seven-LED stripe), a lithium polymer battery in a plastic bag, several tight-knit fabric pieces with floral, blue, green and purple animal print and other patterns for the front cover, a piece of thicker olive fabric for the back side of the device, a piece of lamination foil for textiles, a piece of baking sheet, two pieces of hook and loop fastener strips, a clamp and a loop of 2-ply steel thread in a paper bag. On the right lies a handrawn sign reading “Digi Merkki” and green, blue and pink dots.}
\end{figure} 

\subsubsection{Co-Creation}

\begin{figure}[tbh]
  \centering
  \includegraphics[width=\linewidth]{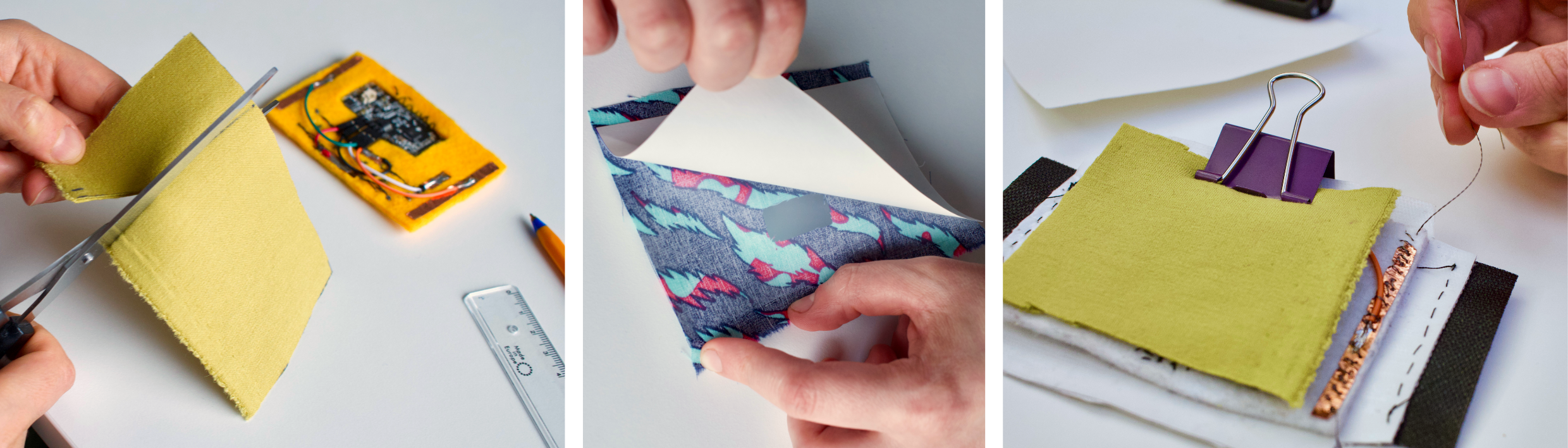}
  \caption{These pictures are examples from the instructions given to the participants.}\label{fig:co-creation}
  \Description{The picture shows three photographs from left to right. The leftmost picture is showing someone's hands cutting a rectangular green piece of thick stretchable fabric on a table where the electronics patch prototype is laying. The central picture is taken after the lamination foil is stuck to the precut fabric piece for the device's cover. It is showing someone's hands removing the paper side of the lamination foil. The rightmost picture is showing how to sew the conductive thread. On a white surface are layered, one of top of the other, the following items: a piece of tight-knit fabric for the cover (only the edges depicted),the electronics patch prototype (only the sides covered with a copper strap are shown), a rectangular piece of thick stretchable fabric. The three layers are hold together by a clamp. Someone's hand is sewing the conductive steel thread through the copper strap present on the patch prototype layer.}
\end{figure}

For building their personalised Digi Merkki, each participant received a creation kit containing the electronics patch prototype, various textiles and a battery (see Figure~\ref{fig:kit}). Participants received detailed instructions for the process (see Figure~\ref{fig:co-creation}) and were free to build their Digi Merkki by themselves or with our assistance during remote workshops. To empower our participants, we provided means to modify and appropriate Digi Merkki throughout the \study{intervention}. The hardware specifications and software are available as open-source\footnote{Felix A. Epp. 2022. eppfel/DigiMerkki: Field Evaluation Prototype. Zenodo. DOI:\url{https://doi.org/10.5281/zenodo.6327712}
}. Further, we encouraged participants to use and modify their patches freely.

\subsubsection{Elicitation Diaries}

After participants had finished their patches, we met them again in personal calls. At his point between the two phases of the study, four participants dropped out of the study. These participants voiced time commitment as the reason for not attending the rest of the study. Hence, we only gathered feedback on the co-creation experience and briefed the remaining 16 participants on the procedure of the \study{field study}.

For capturing the participants’ utilisation of Digi Merkki, we used an event-contingent survey to fill an elicitation diary \cite{jonathanlazarResearchMethodsHumanComputer2010}. The survey contained two parts, “suit-up” and “return”. Participants filled the suit-up survey every time they dressed up and wore Digi Merkki. Participants filled out the return survey when they were on their way home from an event or started charging Digi Merkki. The quick surveys asked for contextualising information, like social ties and activities, and samples of experiences with Digi Merkki. These snapshot diaries are not descriptive but elicit deeper inquiry in the consecutive individual interviews. This shorter process removed the burden of filling detailed diaries while still capturing accounts of subjective experience.  

\subsubsection{Gatherings and Observations}

With a small-scale study in a large student population, the chances of serendipitous encounters between participants were low, mainly because events were limited in size due to the ongoing COVID-19 pandemic. Therefore we staged four small events during the 14 days of data collection. Each event was advertised as an after-work hang-out and provided light catering. The third and fourth events included a barbecue by one of the participants. All gatherings were open ended and lasted between 90 and 120 minutes, and two researchers were attending for participant observations. The two researchers reflected on their observational notes after each gathering.
Additionally, we introduced a competition for collecting the most pictures as an incentive for the participants to engage with each other.

\subsubsection{Individual Post-\study{intervention} Interviews}

People present themselves based on their audience. Hence, it was paramount to interview each participant individually to allow a more intimate reflection on their motives during the social engagements with Digi Merkki. We interviewed all participants of the second phase ($n = 16$) individually during the two weeks after the \study{field intervention}. Participants handed over their patches to retrieve the internal log data. We interviewed them about their experience with Digi Merkki and other participants. Each interview contained individual questions based on the observations and diary entries. The interviews lasted one hour on average (from 43~min to 1:39~h) with a total of 16.25 hours of interview material.

\subsection{Data Collection and Analysis}

We used a mix of collection methods to capture how students perceived Digi Merkki and how participants utilised Digi Merkki as part of  “wearing student uniforms”. The primary source for answering our research questions were the individual interviews. Our analysis drew from:
\begin{itemize}
\item Questionnaire answers on cultural practices
\item Resulting designs of the co-creation (patches and choices of digital pictures)
\item Elicitation Diaries
\item Field notes from the participant observation during four organised gatherings
\item Logging of device interactions of each participant
\item Audio-recordings of the in-depth individualised interviews 
\end{itemize}

%\subsubsection{Reflective Thematic Analysis}

As presented before, this work aimed to investigate participants’ lived experiences, views, and practices. Therefore we chose reflective thematic analysis~\cite{braunOneSizeFits2020}. Informed by the understanding of self-presentation practices~\cite{goffmanPresentationSelfEveryday1959} and socio-cultural practices around fashion~\cite{wilsonAdornedDreamsFashion2003,entwistleFashionedBodyFashion2015}, two researchers coded each half of the interview transcripts. The research questions guide the development of higher-level codes in iterations of reflection between the two researchers. Here, we took a perspective that assumes practices as non-static and identities and behaviour as relative. We then used visualisations of the device interaction data to reflect on our themes. Our analysis generated 25 sub-themes that formed three main themes.

\section{Results}

The following section reports on the three themes we developed in our thematic analysis. We have to note that data points outside the gatherings we facilitated were marginal. Participants reported a lack of student events due to ongoing COVID-19 limitations.
Because students wear boiler suits primarily for group settings, there was little need to wear them and, therefore, little use for the patch.
Nevertheless, the participants described the organised gatherings comparable to other student events. Further, our themes show clearly how Digi Merkki sparked stark responses from our participants and how they adopted and sustained emergent practices. While each section represents a theme, the subsections reference popular memes to better immerse in the studied culture.

\subsection{Digi Merkki Adds to Student Social Practices}

Our first theme demonstrates Digi Merkki as a material in the students’ practices. Here Digi Merkki functioned as an item for adornment and its characteristics brought a new intensity to the practice.

\begin{figure*}[t]
    \centering
    \begin{subfigure}[b]{0.15\textwidth}
        \centering
        \includegraphics[width=\textwidth]{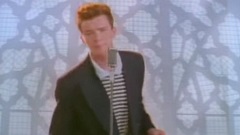}
        \caption{01: -, -}
        \label{fig:merkki_pic_01}
        \Description{Rick Astley in his viral video “Never gonna give you up”}
    \end{subfigure}
    \hfill
    \begin{subfigure}[b]{0.15\textwidth}
        \centering
        \includegraphics[width=\textwidth]{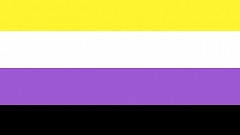}
        \caption{03: -, -}
        \label{fig:merkki_pic_03}
        \Description{Flag of non-binary people}
    \end{subfigure}
    \hfill
    \begin{subfigure}[b]{0.15\textwidth}
        \centering
        \includegraphics[width=\textwidth]{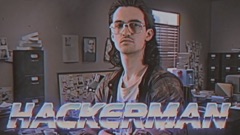}
        \caption{04: 24, 10}
        \label{fig:merkki_pic_04}
        \Description{A “nerdish” looking man in an office with the text written “Hackerman” at the bottom.}
    \end{subfigure}
    \hfill
    \begin{subfigure}[b]{0.15\textwidth}
        \centering
        \includegraphics[width=\textwidth]{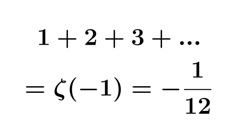}
        \caption{05: 5, 5}
        \label{fig:merkki_pic_05}
        \Description{A math formula.}
    \end{subfigure}
    \hfill
    \begin{subfigure}[b]{0.15\textwidth}
        \centering
        \includegraphics[width=\textwidth]{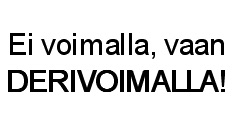}
        \caption{06: 11, 7}
        \label{fig:merkki_pic_06}
        \Description{Black letters saying “Ei voimalla, vaan DERIVOIMALLA”}
    \end{subfigure}
    \hfill
    \begin{subfigure}[b]{0.15\textwidth}
        \centering
        \includegraphics[width=\textwidth]{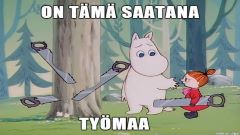}
        \caption{16: 21, 12}
        \label{fig:merkki_pic_16}
        \Description{Moomin comic showing multiple saw stuck in a tree and two Moomin characters next to it. White letters say “On tämä saatana” on top and “työmaa” at the bottom.}
    \end{subfigure}
    \hfill
    \begin{subfigure}[b]{0.15\textwidth}
        \centering
        \includegraphics[width=\textwidth]{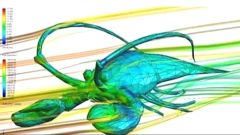}
        \caption{17: 7, 7}
        \label{fig:merkki_pic_17}
        \Description{Screenshot of a lobster in a wind simulation software. }
    \end{subfigure}
    \hfill
    \begin{subfigure}[b]{0.15\textwidth}
        \centering
        \includegraphics[width=\textwidth]{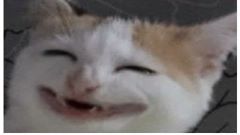}
        \caption{19: 30, 11}
        \label{fig:merkki_pic_19}
        \Description{A laughing cat.}
    \end{subfigure}
    \hfill
    \begin{subfigure}[b]{0.15\textwidth}
        \centering
        \includegraphics[width=\textwidth]{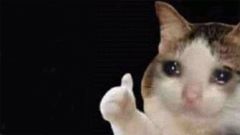}
        \caption{20: 58, 14}
        \label{fig:merkki_pic_20}
        \Description{A crying cat the has a thumb raised up.}
    \end{subfigure}
    \hfill
    \begin{subfigure}[b]{0.15\textwidth}
        \centering
        \includegraphics[width=\textwidth]{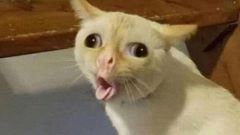}
        \caption{21: -, -}
        \label{fig:merkki_pic_21}
        \Description{A cat that forms and o with her mouse while sticking out the tongue.}
    \end{subfigure}
    \hfill
    \begin{subfigure}[b]{0.15\textwidth}
        \centering
        \includegraphics[width=\textwidth]{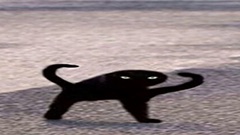}
        \caption{22: 3, 2}
        \label{fig:merkki_pic_22}
        \Description{A black figure.}
    \end{subfigure}
    \hfill
    \begin{subfigure}[b]{0.15\textwidth}
        \centering
        \includegraphics[width=\textwidth]{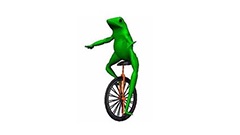}
        \caption{26: 6, 5}
        \label{fig:merkki_pic_26}
        \Description{A frog riding a unicycle}
    \end{subfigure}
    \hfill
    \begin{subfigure}[b]{0.15\textwidth}
        \centering
        \includegraphics[width=\textwidth]{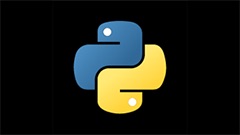}
        \caption{27: 4, 3}
        \label{fig:merkki_pic_27}
        \Description{The python programming language logo}
    \end{subfigure}
    \hfill
    \begin{subfigure}[b]{0.15\textwidth}
        \centering
        \includegraphics[width=\textwidth]{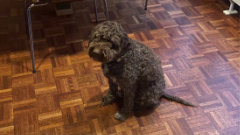}
        \caption{28: 12, 8}
        \label{fig:merkki_pic_28}
        \Description{A dog sitting on wooden floor.}
    \end{subfigure}
    \hfill
    \begin{subfigure}[b]{0.15\textwidth}
        \centering
        \includegraphics[width=\textwidth]{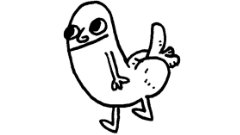}
        \caption{30: 26, 8}
        \label{fig:merkki_pic_30}
        \Description{Dick Butt character}
    \end{subfigure}
    \hfill
    \begin{subfigure}[b]{0.15\textwidth}
        \centering
        \includegraphics[width=\textwidth]{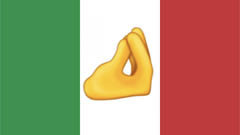}
        \caption{32: 1, 1}
        \label{fig:merkki_pic_32}
        \Description{The Italian flag with the pinched fingers emoji}
    \end{subfigure}
    \hfill
    \begin{subfigure}[b]{0.15\textwidth}
        \centering
        \includegraphics[width=\textwidth]{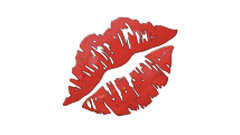}
        \caption{33: 0, 0}
        \label{fig:merkki_pic_33}
        \Description{A kiss mouth}
    \end{subfigure}
    \hfill
    \begin{subfigure}[b]{0.15\textwidth}
        \centering
        \includegraphics[width=\textwidth]{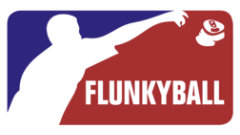}
        \caption{38: 0, 0}
        \label{fig:merkki_pic_38}
        \Description{The baseball league logo adapted to “Flunkyball”}
    \end{subfigure}
    \hfill
    \begin{subfigure}[b]{0.15\textwidth}
        \centering
        \includegraphics[width=\textwidth]{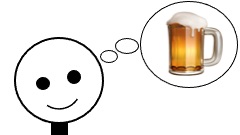}
        \caption{39: 12, 5}
        \label{fig:merkki_pic_39}
        \Description{A stick figure thinking of the beer emoji}
    \end{subfigure}
    \hfill
    \begin{subfigure}[b]{0.15\textwidth}
        \centering
        \includegraphics[width=\textwidth]{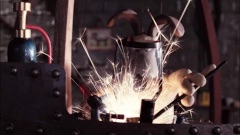}
        \caption{40: 3, 3}
        \label{fig:merkki_pic_40}
        \Description{A screenshot from Wallace and Gromit with Gromit welding.}
    \end{subfigure}
    \hfill
    \begin{subfigure}[b]{0.15\textwidth}
        \centering
        \includegraphics[width=\textwidth]{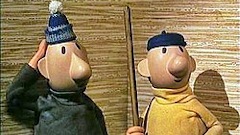}
        \caption{41: 21, 11}
        \label{fig:merkki_pic_41}
        \Description{A screenshot of Pat and Mat}
    \end{subfigure}
    \hfill
    \begin{subfigure}[b]{0.15\textwidth}
        \centering
        \includegraphics[width=\textwidth]{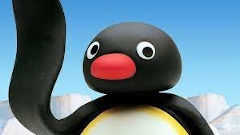}
        \caption{42: 15, 8}
        \label{fig:merkki_pic_42}
        \Description{Screenshot of the Pingu charachter}
    \end{subfigure}
    \hfill
    \begin{subfigure}[b]{0.15\textwidth}
        \centering
        \includegraphics[width=\textwidth]{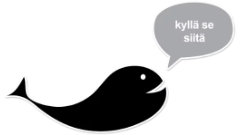}
        \caption{46: 13, 8}
        \label{fig:merkki_pic_46}
        \Description{A whale saying “kyllä se siitä”}
    \end{subfigure}
    \hfill
    \begin{subfigure}[b]{0.15\textwidth}
        \centering
        \includegraphics[width=\textwidth]{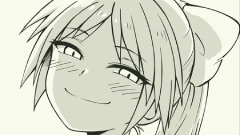}
        \caption{53: -, -}
        \label{fig:merkki_pic_53}
        \Description{A smirking anime character}
    \end{subfigure}
    \caption{A subset of the pictures participants chose. The caption reads as: [identifier]: [times exchanged between participants], [number of distinct participants that received the picture]. The pictures \subref{fig:merkki_pic_01}, \subref{fig:merkki_pic_03} and \subref{fig:merkki_pic_53} were from participants that did not participate in picture trading. Copyright held by \subref{fig:merkki_pic_01}: Rick Astley / BMG, \subref{fig:merkki_pic_04}: Laser Unicorns Productions, \subref{fig:merkki_pic_16}: Moomin Characters Oy, \subref{fig:merkki_pic_30}: KC Green, \subref{fig:merkki_pic_40}: Aardman Animations Ltd., \subref{fig:merkki_pic_41}: Oksana Rogožina / Jan Smrčka, \subref{fig:merkki_pic_42}: JOKER / NHK / NEP / PPI, \subref{fig:merkki_pic_53}: paxiti}
    \label{fig:merkki_pics}
\end{figure*}

\subsubsection[I Can Has Picture?]{I Can Has Picture? \footnote{References “LOLcats”, in particular “Happy Cat” \url{https://knowyourmeme.com/memes/happy-cat}}}

The primary motivation to participate in our study and to engage with Digi Merkki came from the underlying practice of adorning the boiler suit: P7 “The whole [boiler suit] culture is about showing off all the badges you have.” Moreover, Digi Merkki’s lights and screen stand out, especially in nighttime events. Hence, in two accounts of participants using the patch outside our organised gatherings were “an opportunity to show off”(P10). Concurrently, Digi Merkki offered a way to adorn the boiler suit with lights to attract attention and copious visual depictions. Some students aimed for high quantities of pictures for their digital patch, while others looked to collect only particular pictures they were interested in.

Adorning the boiler suits also serves to express the self. Consequently, many participants choose their initial three pictures (see Figure \ref{fig:merkki_pics}) to express their interests (e.g. popular media in \ref{fig:merkki_pics}\subref{fig:merkki_pic_04},\subref{fig:merkki_pic_16},\subref{fig:merkki_pic_41},\subref{fig:merkki_pic_42} or professional topics in \ref{fig:merkki_pic_05},\subref{fig:merkki_pic_27}), affiliation (country in \ref{fig:merkki_pic_32} or student association (not depicted)) or identity (e.g. gender in \ref{fig:merkki_pic_03}). Especially interesting was that picture memes seemed a valuable source to communicate a mood non-verbally: P7: “[Picture \ref{fig:merkki_pic_20}] is my totem animal for the last half a year”. For their picture choices, students often relied on the prevalence of customised stickers in messaging apps, like Telegram, Instagram or Snapchat.

Throughout our \study{field intervention}, participants appropriated Digi Merkki for their own needs. According to the culture, students need to stand out with something unique. This uniqueness showed in the participants’ deliberate design choices. Many participants created some customisation outside of the predefined options. For example, participant three chose a reflective outer fabric, participant ten sewed a brim around the screen (see Figure~\ref{fig:creations-teaser} left), participant five skipped cutting the holes and opted to let the LEDs shimmer through the fabric, and participant 14 integrated some detailed embroidery (see Figure~\ref{fig:creations-teaser} bottom). Three participants went even so far as to construct their own hard- or software. Participant 15 built a handheld device with physical buttons in order to have a more long-lasting, safer device. At the same time, participant 13 developed software to draw arbitrary graphics on the micro controllers screen using a mobile phone through a web application. However, this also removed the original software and P13’s possibility to participate in any of the group interactions provided by Digi Merkki. On the contrary P15 asked to reset his device to the original software to participate, although they had implemented some games for the device.

Another use of Digi Merkki that stemmed from existing student culture practices was reminiscence. Many overall patches relate to past events, and their boiler suit becomes a collection of memories with time. While this presents social status, it also serves students a more inward function. The patches help to remember fond memories and materialise a personal history. This practice showed in the utilisation of Digi Merkki in two ways. First, participants selected pictures for their sentimental value, e.g. as “fun memories from my childhood”(P15). Second, the activity of trading pictures through Digi Merkki brought joy because it resembled Pokémon and other trading games. Finally, multiple participants reported, how they browsed their digital pictures to remember the attached memories.

\begin{figure*}[t]
    \centering
    \begin{subfigure}[b]{0.4\textwidth}
        \includegraphics[width=\textwidth]{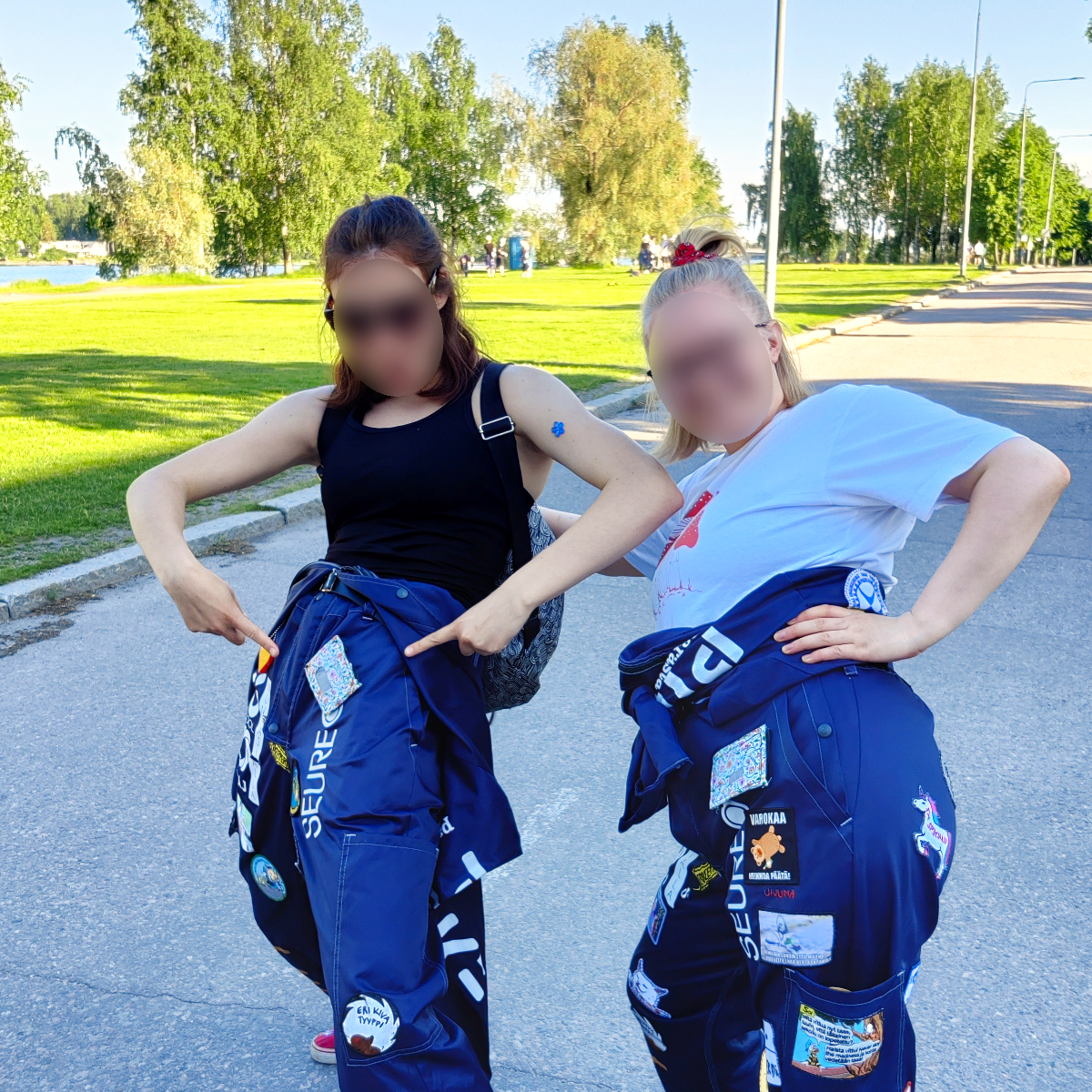}
        \caption{Two participants posing with the Digi Merkki.}
        \label{fig:pose}
        \Description{A photograph showing two persons outdoors on a road with trees in the back. The persons wear blue boiler suits hanging from the waist. The person on the right points with both indexed fingers to textile patch with floral patterns on their waist.}
    \end{subfigure}
    \hfill
    \begin{subfigure}[b]{0.533\textwidth}
        \includegraphics[width=\textwidth]{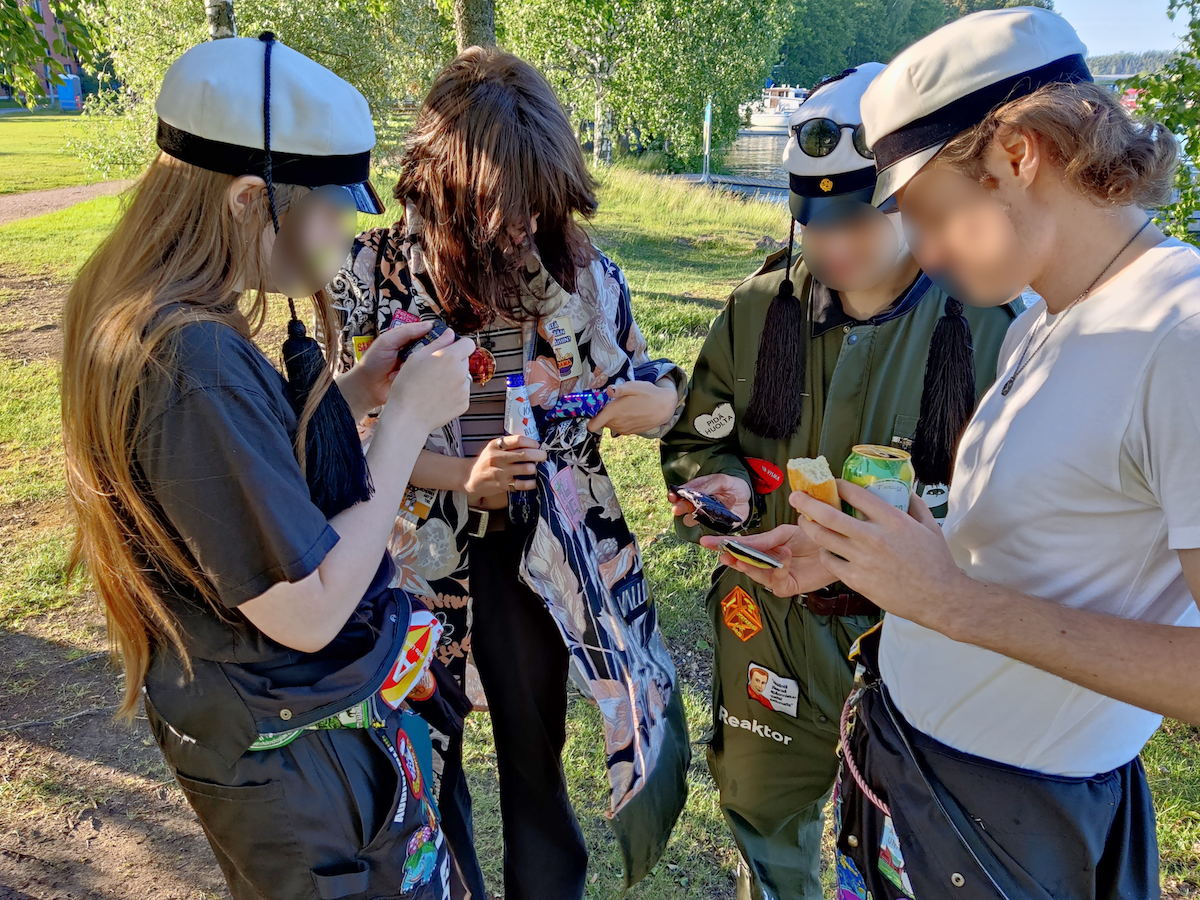}
        \caption{Participants using the Digi Merkki during an event.}
        \label{fig:group}
        \Description{Four students stand in a circle in a green park at a shore facing devices in their hand. Three students wear white caps with short black brim and black tassel. The two people in the front wear black boiler suits until the waist and t-shirts. One person in the back wears a colourful jacket the other a dark green boiler suit.}
    \end{subfigure}
    \caption{Students were wearing Digi Merrki during the diary study (a) and using it actively during the organised events (b).}
    \label{fig:field}
\end{figure*}

In summary, we can note that boiler suit practices revolve around membership by connecting students to their subject associations and the student community as a whole. Students confirmed this explicating their motivation to meet new people in our gatherings. Digi Merkki enabled and facilitated social interaction like a “social lubricant” (P10). Wearing Digi Merkki functioned as a ticket-to-talk~\cite{sacksLecturesConversationVolumes1995}. For example, participant seven joined a board meeting of their subject association wearing Digi Merkki and was approached because of the characteristics of the device: “people were asking me ‘oh, was it these super expensive kits?’” Furthermore, the rainbow coloured proximity-based animations became “a social enabler”(P8). Additionally, wearing Digi Merkki already symbolised some common interests to others, which can function as a ticket.
Lastly, picture trading served as “ice-breakers” (P3) or “conversation starter” (P1). Although pictures were often generic depictions, most pictures bore some personal connections to the wearer, which functioned as a gateway to engage in meaningful conversations. 

\subsubsection[One Does Not Simply Walk Into Conversation]{One Does Not Simply Walk Into Conversation \footnote{References “One Does Not Simply Walk Into Mordor” \url{https://knowyourmeme.com/memes/one-does-not-simply-walk-into-mordor}}}

All these behaviours exemplify how Digi Merkki integrated into the existing social practices of the students. However, Digi Merkki was able to augment these practices with a new intensity, as P13 described: “[Digi Merkki is] both for you as a relic, but also for [signalling to others] …, but it’s stronger in both directions.”
For one, \textit{versatility} of a digital screen allowed people to display rare, missing or forbidden pictures, as P14 explains: “I was interested in getting the Moomin\footnote{Moomins are a popular franchise in Finland} pictures because we cannot have them … on the overall anymore”. Further, this \textit{versatility} enabled participants to choose what they wanted to depict more flexibly at any given point in time. Another characteristic is \textit{expressivity}. It helped a wearer to draw attention to themselves in ways fabric patches cannot.
%[Memes as a form of expressions that suit digital formats even in clothing?]
Lastly, the characteristic of \textit{connectedness} allowed trading pictures rapidly and introduced a non-verbal communication channel. This showed in the affordance of Digi Merkki that led to episodes when participants focused only on their devices (see Figure~\ref{fig:group}), however “were still all together despite not looking or talking to each other” (P8).
This was in contrast to participant 13 and their customised prototype. Participant 14 viewed that prototype problematic, as it set a focus on phone interaction over Digi Merkki: “If [they are] always handling the [interactive patch] over [their] phone … then [they] are just on [their] phone. … people will not notice it as what it is and it will probably not create the interesting conversation.”

\subsection{Digi Merkki Fosters Emergent Social Practices}

In the second theme we describe how participants adapted new ways of interacting with Digi Merkki and each other. They appropriated Digi Merkki in their own ways to introduce dare challenges and playfully teasing each other. As knowledge of such novel usage spread between participants, emergent social practices were formed in this particular group.

\subsubsection[Challenge Accepted]{Challenge Accepted \footnote{References “Challenge Accepted” \url{https://knowyourmeme.com/memes/challenge-accepted}}}

A group of three participants attached a dare challenge to each of their pictures before the \study{intervention} started. To receive one of their pictures, a trading partner first had to fulfil the respective challenge.
For example, participant ten required a kiss for picture 33 and picture 38 was a trophy for beating participant two in a drinking game (see Figure~\ref{fig:merkki_pic_39}).
While this practice of pictures as achievements added game-like character to picture trading, participant five described the challenges as “one way of giving [the pictures] meaning”.
The three participants each purposefully thought of one easy challenge to use as a conversation starter.

\begin{figure*}[t]
  \centering
  \includegraphics[width=\textwidth]{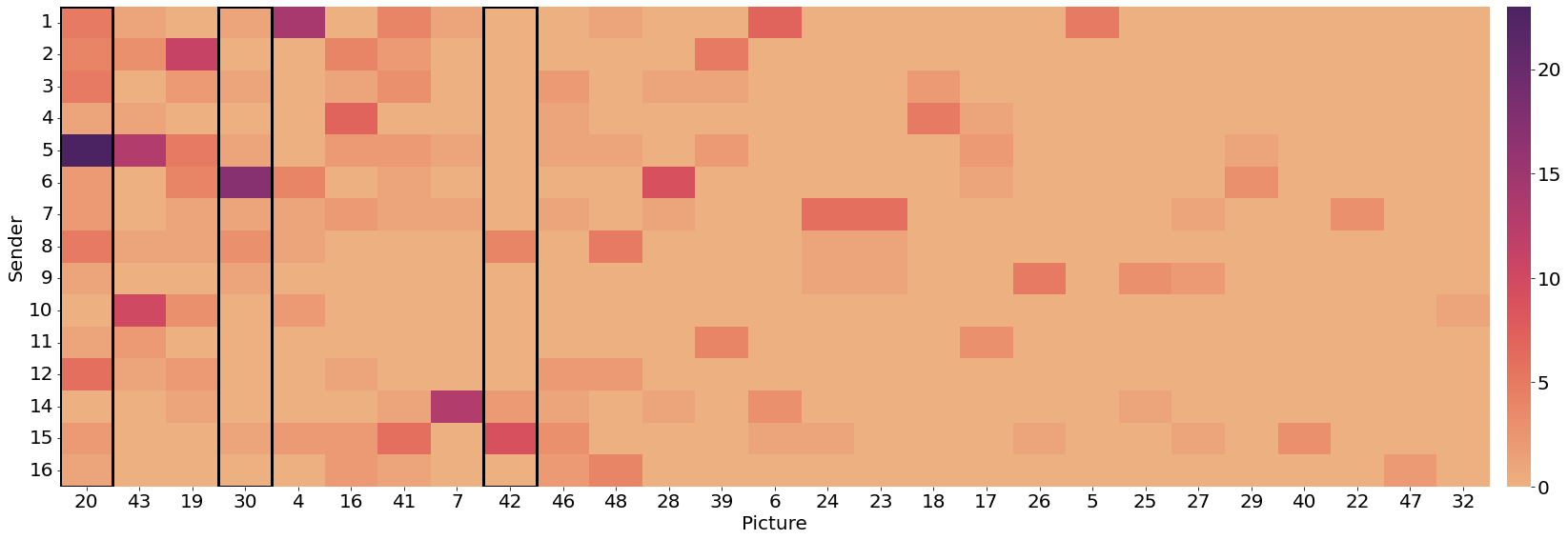}
  \caption{Heatmap depicting how often a participant shared a particular picture. The high-intensity (i.e., dark red) spots clearly show participants chose specific pictures to send repeatedly. The framed columns represent the pictures in Figure~\ref{fig:pic_sharing_plots}. Pictures are ordered from highest to lowest sharing.}\label{fig:heatmap}
  \Description{A heat-map chart with one axis with 15 senders and one axis with 25 pictures showing the number of sends from zero to 25 for each section of the raster. The chart is ordered from left to right with the highest to lowest shares. The top five picture columns show high intensity, while the lower half shows almost none. In particular, single cells in various picture columns show extremes.}
\end{figure*}

In the course of the \study{field intervention}, these dare challenges were not only upheld by their creators but also adopted by other participants. By the end of our data collection, each of the three initiators (P2, P5, P10) had still not shared one picture because no one had fulfilled its challenge (see Figure~\ref{fig:merkki_pics}(33,38)). Some other participants that encountered the challenges adapted such “rules” (P1, P14) for their pictures. The challenges spread over multiple participants, as participant one adopted them from participant 14, without knowing another group (P2, P5, P10) had initially introduced them. 

Another form of challenge developed from the prevalence of certain memetic picture sets. Multiple participants had picked similar pictures, e.g. cats or Moomins. This motivated participants to collect these particular pictures as a set. In conclusion, participants found ways to induce meaning, whether through scarcity or achievements.

%P1: “scarecity brings value.”
%P5: “I have achievements for my pictures, my pictures mean something, my pictures hold value.”

\subsubsection[The Struggle is Real Fun]{The Struggle is Real Fun \footnote{References “The Struggle is Real” \url{https://knowyourmeme.com/memes/the-struggle-is-real}}}

The emergent social practice which changed Digi Merkki’s reception and utilisation the most was a form of spamming.
Due to the time-based picture trading interaction, a third player could spam the network by constantly holding the respective seam for sharing. Once any other participant started a trade, they consequently traded pictures with the spammer. This aspect was neither intended in the design nor had it been discovered in the pilot study. Participants coined this as “spamming”, “stealing”, “snitching”, or “hi-jacking”.
While the act of spamming interrupted one to one social interactions, it was seen by participants as a way to enable group interaction.
In particular, participant 16 used spamming to enter a dyadic interaction from the outside: “[P12 and P13] were exchanging some pictures, so I just sneaked in there.”
This act of spamming was seen as “friendly teasing” (P1), “annoying in a fun way”(P8) or “act[ing] mad because it’s fun.”(P7).
These reactions seemed to stem from the openness of the design, as participant eight formulated aptly: “Since it wasn’t necessarily enforced [to] trade with the person in front of you, there were no rules …, it made the hijacking possible, so acceptable and funny.”
The spamming added a surprise element to the trading activity. Participants liked the its chaotic nature of receiving unexpected pictures. This unexpectedness affected the people that tried to trade, but also spammers: “Because it’s not guaranteed that I’m able to snatch it away. If it works, it’s still a small achievement”(P5). This heightened feeling of achievement was based on the uncertainty of the interactions and introduced another game-like aspect to the social interactions. By actively engaging in spamming, the participants made picture trading more effortful and, consequently, certain pictures more meaningful.

%\subsubsection[How does the cat say?]{How does the cat say?}%\footnote{ \url{}}}

\begin{figure*}[t]
    \centering
    \includegraphics[width=\textwidth]{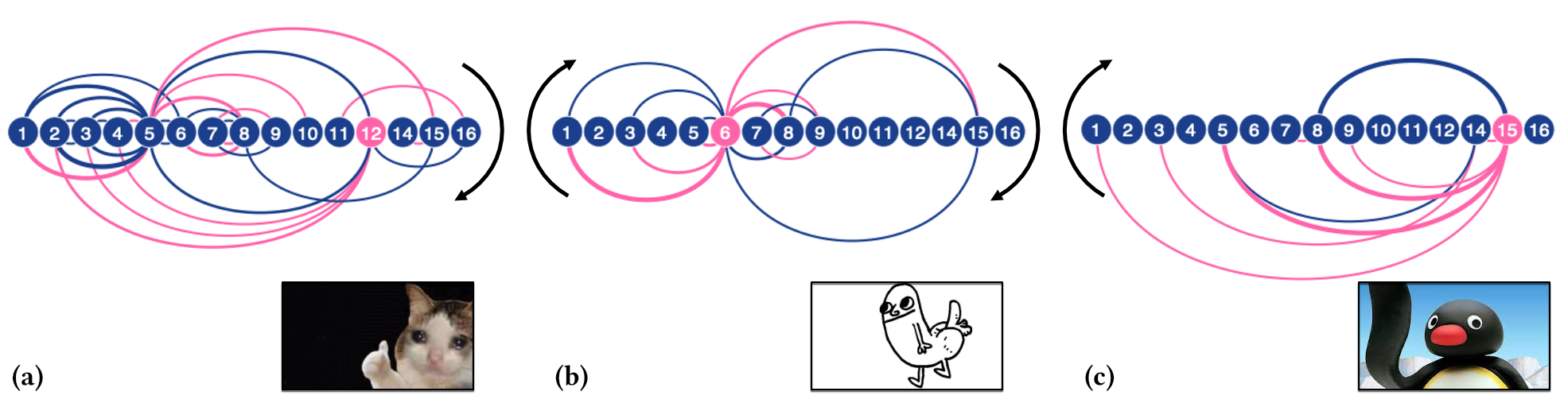}
        \caption{Network graphs showing how 15 participants shared the particular pictures a:20, b:30 and c:42. Nodes represent participants and lines show unidirectional sharing between two participants in clock-wise direction. The thickness corresponds to the amount of times a picture was sent. Nodes in magenta represent the initial owners of the pictures and the magenta arrows show how the picture initially spread through the network. Blue arrows show the reception of pictures once that picture was already in the receivers possession. Thumbnails copyright: b: KC Green, c: JOKER / NHK / NEP / PPI}
        \label{fig:pic_sharing_plots}
        \Description{Three network graphs from left to right a,b,c showing how 15 participants shared the thumbs up cat picture no. 20,  the dick butt picture no. 30 and the Pingu picture 42. In each graph are 15 nodes representing participants and lines show unidirectional sharing between two participants in clock-wise direction. Various thickness corresponds to the amount of times a picture was sent. In the left graph node 12, in the centre graph node 6 and in the right graph node 15 are in magenta and represent the initial owners of the pictures. While a lot of magenta lines in the graph b and c show how the picture initially spread through the network to a lot people, in graph a almost as much blue lines show the reception of picture often once that picture was already in the receivers possession. In graph a most lines go out from participant 5, while in the other graphs from the respective owners.}
\end{figure*}

The participants even used dedicated pictures to communicate the teasing. As you can see from Figure~\ref{fig:pic_sharing_plots}a, in particular, participant five sent the  “thumbs up cat” again and again to all other participants. Other dedicated spamming pictures were “hackerman”(\ref{fig:merkki_pic_04}), “laughing cat”(\ref{fig:merkki_pic_19}) and “dick butt”(\ref{fig:merkki_pic_30}), which shows in the Figure~\ref{fig:heatmap} as high-intensity spots. In contrast, the picture showing “Pingu” in Figure~\ref{fig:pic_sharing_plots}c was also popular but not reshared after people had it in their collection. Hence, participants reappropriated the cat picture (\ref{fig:merkki_pic_20}) of P12 to communicate the humouristic nature of teasing. As participant 14 paints it: “[The cat] is telling you like ‘haha, don’t do this you’ll anyway just get me again’”. For this reappropriation of meaning, the picture memes seemed particularly useful, as all the most shared pictures have some meme-like characteristics (see Figure~\ref{fig:heatmap}).

\subsection{Students Navigate Divergent Emerging Practices}

Finally, our third theme highlights how participants actively shaped the practices around Digi Merkki based on their individual competences. Participants at times had contrary aims for using the spamming or the Digi Merkki as an ice-breaker. However, they formulate clear proposals how to navigate and even shape those through building structural elements in the community or their own agency.

\subsubsection[Well, That Escalated Quickly]{Well, That Escalated Quickly \footnote{References “That Escalated Quickly” \url{https://knowyourmeme.com/memes/that-escalated-quickly}}} \label{sec:results-escalate}

While some participants attached importance to contributing to the community, others were more interested in propagating and promoting an idea. Many participants thought carefully about what pictures might be interesting for others. However, some participants picked pictures that only represented their interests. For example, participant one intended to spread their love for mathematics, hoping that more people would get interested in this topic. This way, they would have more people with a common interest. This strategy also showed in the way they appropriated the challenges. While participant five wanted to increase the value of their pictures by attaching achievements to them, participant one adapted the challenges differently. They recognised how challenges added meaning to the pictures and used this again to promote mathematics through the maths meme (see Figure~\ref{fig:merkki_pic_05}). Furthermore, participant six made it their personal challenge to spread their “dick butt” picture even against others interest: “[they] really did not like the picture. I wanted to have that picture sent to [them] in a playful way, nothing malicious or anything.” As apparent from the graph in Figure~\ref{fig:pic_sharing_plots}b dick butt mainly was sent from participant six to others, while others rarely send it forward. Participant 13 even aimed for controversy with his choice of images: “I want to propagate the weeb shit.” They hoped to make their ideas part of the mainstream: “It becomes more of a normy thing”.

While most people enjoyed the competition as something “not too serious”, there were also divergent practices in trading pictures. On one side, participant 15 handed their device over to others, so they could go through the pictures and take the ones they preferred. For them, “the main idea was that we would eventually get all the pictures”. On the other side, some saw the spamming exploit more as a chance for stealing: “I was more thinking of getting pictures for myself rather than letting people have mine”(P10).

\subsubsection[If we could stop stealing pictures, that would be great.]{If we could stop stealing pictures, that’d be great. \footnote{References “That Would Be Great” \url{https://knowyourmeme.com/memes/that-would-be-great}}} 

We have reported how Digi Merkki was a conversation starter. Participants that describe themselves as introverts explained how this helped to avoid “awkward” small-talk: “You start the conversation that way, without having to [say] ‘oh what have you been doing this weekend?’”(P1). Meanwhile, participant 13 did not bring their custom device to the first gathering. Therefore, they could not participate: “The point was to use the devices. I didn’t have it, so it was awkward for me”. Of course, the study setting created this distinction because it gave the gatherings a clear context. However, even when participant 13 joined with their custom application, others noted how not participating in the interactions mediated by Digi Merkki excluded P13, 
as participant described it: “[P13] could engage …, but [they] couldn’t understand or relate.… I felt [they] couldn’t be part of the community.”
Hence, participants recognised problems of inclusion and even expressed their wishes for developing such technology further. Participant 14 said, “It would be sad if [Digi Merkki] would be an exclusive thing. I would wish that it could be shared with people all over the place.” They suggested rules for the community to limit the stealing and incentivise asking for pictures. As stated before, participant 15 saw it more relaxed:  “Memes come and go, and if something is funny and makes people laugh, it’s worthwhile.” 

In summary, participants dealt differently with the unprecedented social interactions around the interactive clothing patch. Beyond personal strategies to navigate these group practices, participants discussed and proposed structures to formalise some of these practices. Hence, they took ownership of shaping the practice.

\section{Discussion}

Despite decades of HCI research on social wearables, they have yet to achieve widespread adoption. Thus, this work aimed to gain a better understanding of what drives and hinders adoption of social wearables by studying how they shape adornment and social practices \emph{in-situ}. In the following, we discuss the implications of our work for the understanding and design of social wearables in particular, and social technology more generally.
%Specifically, our field study set out to explore how Nordic students adopt a social wearables in their practice around adornment of boiler suits.

% 1. the intervention! became part of everyday practices
% 2. it lead to emergent practices
% 3. those however were manged by the community !!! -> mechanism of negotiating meanings, identity, but also norms and consequentially acceptability
% 4. this means design should suppoort this self-regulating processes
% 5. how did digi merkki do this: participatory process aiming for involvement
% 6. these characteristics are memes!

% 1.	The study was not evaluating interactions with Digi Merkki but looking at the effects of our design intervention (co-creation with the kits, planning those gatherings with incentives and change of practices with Digi Merkki as a new element in their practices).
% 2.	That a perspective on dialogue and negotiation of meanings (what is private? What is acceptable?, etc) should help adoption and technology can mediate this dialogue and negotiation, as our design intervention showed.

\subsection{Varying Adoption of a Social Wearable Reveal Negotiation Practices}

\tpc{Both in the wearable co-creation process and the field study, we observed the students’ utilisation of Digi Merkki and the emergent patterns in their social practices.}
As noted, the adornment practices of Nordic students revolve around developing identity through distinction and group membership.
We saw how our design intervention contributed to the practice of meeting new people as a conversation starter (“social lubricant”) and adding a digital component to their boiler suit adornment.
Furthermore, the activities during the intervention and the collected pictures contributed to reminiscence, just like analogue patches and other student events.
Additionally, we observed emerging patterns with regards to students negotiating individuality and membership in the form of dare challenges and spamming.
We also saw how our participants made use of the ambiguity~\cite{Devendorf2016} Digi Merkki provided and were able to perform their social identity~\cite{dunneSocialAspectsWearability2014}.

\tpc{Especially the emergent practices revealed conflicting strategies in peoples utilisation of Digi Merkki.}
Considering our first research question, we saw how some participants introduced dare challenges, but others adapted them to their needs, as this particular practice spread through the group. Some participants saw value in those challenges for making their personal pictures more meaningful, while others used that effect to spread a particular idea for their benefit.
While most enjoyed spamming as a way of friendly teasing, there were accounts of antagonism between people (see the example of Dick Butt in section~\ref{sec:results-escalate}).
However, Digi Merkki also indirectly led to awkwardness and exclusion when one participant could not participate in the networked features. 

\tpc{However, participants navigated these consequential tensions and awkwardness by negotiating meanings.}
When Digi Merkki sparked a change in practice, this required the participants to redefine the meanings constituting their cultural practices. However, meanings, e.g. the contents of a picture or the aesthetics of a garment, are socially defined. Therefore we can use a dialogical perspective of meanings not as fixed elements transmitted between people, but unstable outcomes of ongoing negotiations~\cite[][chps. 6 \& 7]{barnardFashionTheory2014}. 
We observed how meaning of content (i.e., the pictures participants chose) changed. For example, the thumbs-up cat changed to an expression of superiority and consequently became part of expressing social identity. However, this required negotiation between the individual participant and the community. In fact, also the meanings of people’s actions were negotiated. After all, participants affirmed spamming as a fun activity that enabled group interactions, despite it being uncontrollable.

As we have outlined in section~\ref{sec:related-work-designing}, adoption of a product has been explained through the concept of acceptability, i.e. aesthetics, identity, and privacy~\cite{dunneSocialAspectsWearability2014, olschewskiCollaborationTechnologyAdoption2018, dourishRethinkingPrivacy2011}.
The factors influencing adoption are attached to meanings (What is private? What is acceptable? How does a wearable express social identity?). When applying a dialogical perspective, we can view technology as a way to mediate the negotiation of those factors. Therefore, we see potential in designing for a social wearables that mediates the negotiation of meanings, which provides people with the potential to adopt a technology in their lives.
%Dunne and mechanics: a mechanisms to mediate that acceptance process?”~\cite{dunneSocialAspectsWearability2014} 

\subsection{Characteristics Supporting Involvement and the Mediation of Meanings}

\tpc{The characteristics of our intervention encouraged involvement.}
Specifically our participatory process--starting with the co-design, followed by co-creation of the digital patch and creating meetings in the participants environment---increased democratic and bottom-up social organisation.
%After all the whole process was catered towards involvement: 
The co-design of the prototype resulted in an artefact relevant to the students. The creation kit provided means for personal involvement through personalisation. For example, the students could solve the question of acceptable positioning of the patch on their body themselves by sewing on the patch anywhere and attaching it flexibly with the hook and loop fastener. 
Similarly, Digi Merkki had do-it-yourself characteristics, which contributed to its appropriation. 
%As attested by the responses to our \study{field intervention}, the openness of our design invited participants to shape how they would adopt Digi Merkki.
%More particular certain characteristic lead participants to memes as competences to shape their practices
Digi Merkki’s screen allowed full-colour pictures but constrained them into a compact format. Participants also cleverly used pre-existing memes in their pictures, which invited others to use them for their own purpose.
Consequently, the appropriation led to more involvement and made it more meaningful to the participants.
% Participants of Digi Merkki submitted their initial pictures, and by trading with others, they could add to their memetic repertoire. Ownership was socially defined, not through our system.
%Pictures were selected by people themselves

\tpc{Additionally, the loosely defined interplay of actuation and sensing also enabled students to mediate identity and meanings.}
Digi Merkki did not dictate a use scenario as neither the meanings of trading a picture, nor the meaning of content was defined. The functionality relied on the participants to balance fair use between them.
%and \textit{how people use it between themselves} was left open.
This openness allowed for the emergence of spamming and the introduction of dare challenges.
The possibility of the spamming led to fun. Some participants enjoyed the feeling of superiority created by discovering it; others enjoyed the incongruity of the unexpected behaviour. Albeit indirect, we see this as a notion of mischief and humour as a resource for design~\cite{iivariArseingWasFun2020}. Additionally, this made interactions tricky and introduced problems. However, the extra effort made the actual interpersonal interactions more meaningful, which matches with findings in existing work~\cite{kellyDemandingDesignSupporting2017}.
%A social wearable design needs to step out of the magic circle introduced by a game to allow these balancing acts in everyday scenarios. This resonates with research that forced gamification can lead to rejection(SOURCE). Instead of fixed game rules, 
The balance of control that Digi Merkki provided, enabled participants to reappropriate the picture-trading for spamming but still protected participants’ “exclusive” pictures (as two participants never gave away a picture due to their unfulfilled challenges).
In conclusion, the characteristics of our intervention invited for appropriation and personalisation and this way supported mediation of meanings.  
\todo{
%The value of a badge comes from the interactions around, e.g. achieving a high amount of points in a pub crawl. through this patches are meaningful elements in their adornment practices. Similarly, Digi Merkki and its pictures become meaningful through the history attached to it from trading with other students. 
    %Contestable???
        %-> empowering? reference result on ownership of an (new) practice. 
        %-  people navigated, they are fine. however design needs to acknowledge participatory culture
}

\begin{figure*}[t]
  \centering
  \includegraphics[width=0.95\textwidth]{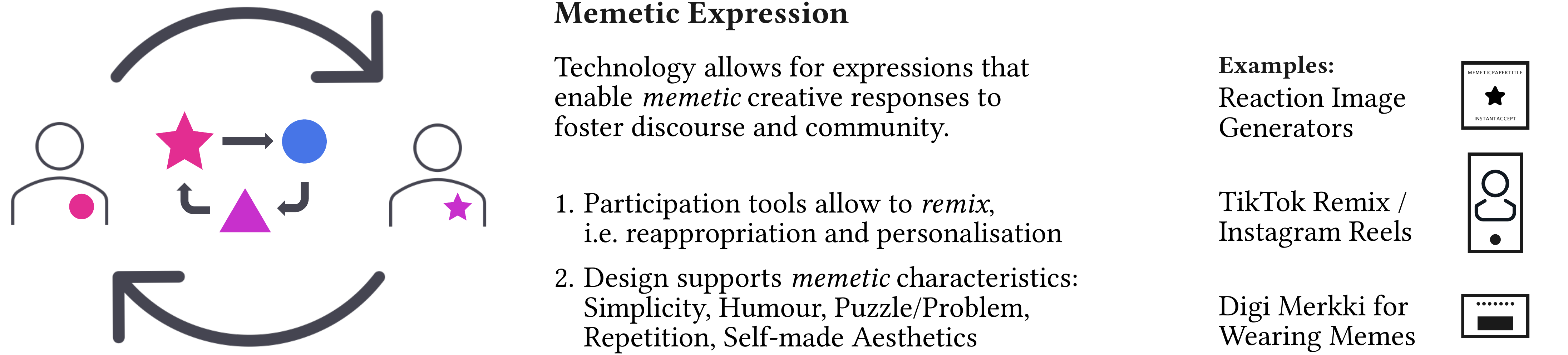}
  \caption{A diagrammatic representation of the strong concept “Memetic Expression”.
  }\label{fig:memetic-expression}
  \Description{
  On the left shows a visualisation of two black arching arrows pointing clock-wise between two people, one with pink circle and one with purple star. In between are a blue circle, a purple square and a pink star each pointing from one to the next in a clock-wise loop.
  In the centre a heading reads "Memetic Expression" with text underneath: “Technology allows for expressions that enable memetic creative responses to foster discourse and community.” Underneath is a list of two principles: 1. Participation tools allow to remix, i.e. reappropriation and personalisation, 2. Design supports memetic characteristics: Simplicity, Humour, Puzzle/Problem, Repetition, Self-made Aesthetics.
  On the right are the examples depicted in abstract black drawings: Reaction Image Generators, TikTok Remix / Instagram Reels and Digi Merkki for Wearing Memes.
}
\end{figure*}

\subsection{Memes in Digital Culture Foster Social Identity and Community}

\tpc{Finally, our work contributes to the design literature on social wearables by identifying that the observed characteristics of our intervention appear to correspond to success criteria of memes in digital culture. }
First, we note that the concept of \emph{internet memes} was not a part of our design process, nor a theoretical starting point for our analysis. Instead they emerged as a theme that cut across all other themes, through the materials participants chose to include into their digital patches, as well as their interactions. 
%The prevalence of internet memes, as pictures for trading and expressing reactions and a structuring concept for the dare challenges, demonstrate their usefulness.
%Memes were helpful to our study participants because they already symbolise a specific context.
%They came in the form of pictures but also challenges.

%Based on our participant’s verbal descriptions, we identified the concept of internet memes, 
As defined by Shifman~\cite{shifmanMemesDigitalCulture2014}, “internet memes” are groups of content units that invite imitation.
In contrast, a \emph{viral} is a single distinct piece of content that is shared rapidly through the internet. Memes, however, are potentially only shared in fringe communities. Following this, we observed participants using memes as elements of meaning and competence in their social practices. Through the memes’ diffusion in their existing practices---i.e. in online communication and boiler suit adornment---the students use the existing meanings of those memes and their competencies in \emph{memeing}\footnote{Memeing. Know Your Meme. Retrieved February 28, 2022 from \url{https://knowyourmeme.com/memes/memeing}
}. Concurrently, fashion studies identified memes as a trend in clothing design~\cite{skjulstadVetementsMemesConnectivity2020}. Here, garments are designed to become memes, that invite for copying and imitation and are not perceived as an “original”. While HCI scholars have recently explored memes as a design material \cite{terzimehicMEMEoriesInternetMemes2021}, they have to date not been discussed as a concept for the design of technology in collocated social interactions.
%\tpc{ competences and meanings of sharing memes was helpful to our study participants as a vessel for communication}

\tpc{These attributes reveal memes as a tool for involvement and for expressing membership to ever smaller groups and building community within them.} In our study, we could see how the practice of adorning boiler suits and the practice of creating and sharing memes intertwined. As previously noted, the student practices around the boiler suits mainly revolve around developing and shaping identity~\cite{eppIdentitySocialWearables2020}: A student’s collection of patches creates a distinction from the “uniform” of the student association, while expressing membership based on the patch’s content. By adding a digital component to adornments, students foremost utilised memes. These memes offered a new granularity of distinction through imitation, which references existing meanings, while at the same time introducing an individual component. These characteristics offer the students an easy tool to build identity and community.
To this end, we articulate a more general abstraction of this mechanism in the form of a strong concept for design. 

\section{Memetic Expression as a Strong Concept for Design}

\tpc{We propose \textit{Memetic Expression} as a strong concept~\cite{hookStrongConceptsIntermediatelevel2012} for designing social technology.} Memes foster discourse and therefore offer an ideal tool for designers to build identity and community. 
The concept of Memetic Expression suggests wearables (or other social technologies) should foster remix and emergent use. Instead of striving for viral success, we focus on the inclusive and participatory aspects of memes, which enable or facilitate social interaction. By leveraging the strong concept of Memetic Expression, we make social wearables an inclusive social participation tool. 

\tpc{Strong concepts \cite{hookStrongConceptsIntermediatelevel2012} embody design knowledge between particular instances and generalised theories. Memetic Expression matches the criteria of a strong concept as it is agnostic towards technology or specific system instances; it emerges through interaction; and it speaks to the behaviour and practices of the users while remaining a part of the design artifact.}
Memetic Expression addresses changing practices at its core. While the concept was popular in our context, it reflects properties of the participatory culture prevalent in the digital age in general. 
Our case demonstrates how Memetic Expression works in different types of social technologies; While memes are primarily found in computer-mediated-communication, the users transferred this concept easily to face-to-face technology, supporting the potential of the concept to cut across into the domain of collocated social interactions. As a strong design concept, Memetic Expression considers especially the layer of sensing and actuation interplay, yet is not restricted to a particular form.

\subsection{Principles}

\begin{table*}
  \caption{Design principles of Memetic Expression as a strong concept}
  \label{tab:memetic-principles}
  %\begin{tabular}{l l }
  \begin{tabularx}{\linewidth}{l X}
    \toprule
    Participation Tools & \textit{Are users actively involved in shaping the sensing--actuation--interplay? What are the tools the design offers for participation? How can users/wearers express themselves freely? Does it support a remix of meanings?} \\
    \midrule
    Simplicity & \textit{Does the digital expression allow for pre-existing symbols? Does the actuation lie on the sweet spot between explicit and implicit information?} \\
    Humour & \textit{Does the design allow for unexpected use and surprises? Does the design build upon playfulness?} \\
    Puzzle/Problem & \textit{How can participants make the interaction more effortful, hence meaningful? How is the design supporting gameful task fulfilment?} \\
    Repetition & \textit{Does the system support repetition by reducing barriers to exchanging (memetic) content? } \\
    Self-made Aesthetics & \textit{What about the aesthetics of the design expresses something self-made? } \\
    \bottomrule
    \end{tabularx}
  %\end{tabular}
\end{table*}

The concept of memetic expression addresses technology that mediates interpersonal communication. The core principle makes use of how memes work. Successful memes are expressions that enable memetic and creative responses. The particularities of memes support identity development and forming group membership. Through these processes, memes, not virals, build community.

Shifman~\cite{shifmanMemesDigitalCulture2014} describes factors that make memes successful: participation tools, simplicity, humour, puzzle/problem and memetic potential (i.e. repetition and self-made aesthetics). We build on these factors and formulate guiding principles and questions for designing social wearables and expressive social technology. The overarching aim  of Memetic Expression is a design that fosters “memetic” creative responses between its users to drive discourse and community. 

\subsubsection{Participation Tools for Remix}

The first and most concrete principle of memetic expression aims to provide users with tools for participation beyond mere propagation. The technology needs to offer involvement that allows people to “reappropriate and personalise universal content”~\cite{shifmanMemesDigitalCulture2014}, in short, “remix”. Remixing content heightens a users sense of ownership and agency. %\todo{Referencing “Produsage” https://en.wikipedia.org/wiki/Produsage Immitation! Anna has a paper on co-creation support!}

\subsubsection{Supporting Memetic Characteristics}

The second principle guides which characteristics of remixing should be supported by a design. These characteristics contribute to the memetic success of content and therefore increase the chances of fostering discourse and community. The characteristics are: Simplicity, Humour, Puzzle/Problem, Repetition and Self-made aesthetics. 
For memetic success, not all characteristics are necessary but a combination increases the chances for propagation. 

\emph{Simplicity} borrows from the memetic success factor of \textit{simple packaging}, which allows content to be shared or imitated easily. Design can support simplicity by introducing constraints. Constraints can support creativity and creative responses heighten a meme’s chances to spread. Further, simple messages are easier to understand quickly. Limiting expressions to specific formats can help people to achieve simplicity. 

\emph{Humour} is a core characteristic of memetic success. In the design of interactive technologies, humour~\cite{iivariArseingWasFun2020} and valance~\cite{colleyExploringPublicWearable2020} have been suggested. While researchers have described frameworks on playfulness (e.g. wearables~\cite{burukDesignFrameworkPlayful2019}), the memetic expression concept does not aim for gaming devices. Instead, it supports social interaction in general. In this way, a design should allow for humorous content without thinking of it as a game.

Successful memes always introduce a \emph{puzzle or problem}, which again invites participation. We know that game-like mechanics foster social interaction~\cite{isbisterInterdependentWearablesPlay2017}. However, again the system must invite for gameful tasks fulfilment and not enforce specific game mechanics. Designers can support this through mechanics indirectly linked to competitiveness. This way, users can introduce their own challenges or competitions. 

Another factor that supports memetic success is \emph{repetition}. When design introduces few consequences for sharing content, the barrier for resharing is lower. Digital designs efficiently cater for lower friction, and designers can directly replace existing analogue expressions to that effect. 

Especially, picture memes often follow typical formats that resemble \emph{self-made aesthetics}.  The appearance of something self-made lowers the barrier for others to modify it. Designers can introduce this by enabling participation tools that make it easier for users to produce something of their own. Designers can aid users by copying existing memetic formats. 

\subsection{Other Examples of Memetic Expression: Image Macros and Mobile Short Videos}

The success of memes as vessels for communication is apparent through their omnipresence in online communities. To visualise the strong concept, we showcase some designs leveraging memetic expression.
While our study aimed at fixing the gap in the adoption of social wearables, there are successful examples of older and more recent designs in mediated communication that show how memetic expression can foster discourse in other domains. 

\emph{Image macros} are considered as the most popular form of picture memes. Several services such as “Meme Generator - Imgflip”~\footnote{“Meme Generator - Imgflip”. Retrieved January 8, 2022 from \url{https://imgflip.com/memegenerator}} offer users tools for creating meme images. These services make use of Memetic Expression contributing to the adoption of image macros on a wide-scale. First, they offer participation tools though collections of meme templates that make it easy to imitate known memes while creating something unique. They also cleverly support memetic characteristics as image macros build upon the concept of a canned joke~\cite{dynelHasSeenImage2016}. Humour is often provided by the meme template. Simplicity is achieved through the standardised format of a reaction image plus two short lines of text. As the reaction text is normally split in to a top and bottom part this pun format enables a puzzle. 

The success of the social media platform TikTok\footnote{TikTok - Make Your Day. Retrieved January 8, 2022 from \url{https://www.tiktok.com/}} can be ascribed to its design being “geared toward imitation and replication”~\cite{zulliExtendingInternetMeme2020} and therefore it corresponds to the concept of memetic expression. Again, the first principle of participation tools is provided by the use of other video and soundtracks as a starting point for a remix and various pre-built filters and effects for customisation. While the content are short videos, which offer a wide variety, the format again is limited in time and orientation. The constraints in format support the characteristics of humour and simplicity. Similar to aspects of successful memes on youtube~\cite{shifmanMemesDigitalCulture2014}, repetition is an often used characteristic on TikTok. Repetition is supported by the easy editing tools that help to quickly generate copies and watching content based on a group of remixed content. The instant messaging service Instagram copied most of these design elements in their “Reel” feature.\footnote{Instagram Reels — Share \& Create Short Videos. Retrieved 1 January 2022, from \url{https://about.instagram.com/features/reels}} And the video platform YouTube, not only added a short video feature in 2020 but also introduced a way to remix existing videos by introducing a “create” button to their player interface\footnote{2021. YouTube Adds a “Create” Button for Shorts in Player Interface. Beebom. Retrieved January 8, 2022 from \url{https://beebom.com/youtube-adds-a-create-button-for-shorts-in-player-interface/}}

\section{Limitations and Future Work}

\tpc{First, }%we have to acknowledge the limitations of our prototypical design.}
while Digi Merkki was adequate as a research prototype, its design is lacking compared to a finished product. The limitations of today’s smart textiles are undoubtedly a hindrance for adoption. In our case, the small resolution of the screen added a useful constraint, but its dimensions also lead participants to hold the device in their hands while interacting, which could have distracted or detracted from engaging with other students. A flexible display would be an adequate solution to support wearability, but those are not yet readily available for rapid prototyping. Furthermore, coated flexible electronics would make the digital patch water-proof to sustain the weather conditions as well as the drinking and bathing practices of the students. Finally, a more refined iteration of Digi Merkki would need to reduce the trading distance of images to a couple of meters to avoid an overload of connections for the devices and the users in larger gatherings. Additionally, Bluetooth or another low power transmission technology would be more suitable for permanent use. %Therfore, efforts in integrating  electronics into existing fabrics are still very much needed.

\tpc{Next, we note that our field study presented a unique case and therefore its results have to be seen in context.} Our study employed a picture trading competition as a means to spark engagement with Digi Merkki. Yet such incentives make it difficult to generalise the effects of the interactive clothing patch. However, we consider the results as an outcome of the intervention. When designers want to consider social practices with wearables, they are required to look beyond the individual moment-to-moment device interactions. In the case of Digi Merkki, the practices of collecting high quantities of patches and competitions to earn rewards are already part of Nordic student culture (see section \ref{sec:case}). 
Nevertheless, the incentive on trading pictures might have led to less reporting on the proximity feature. Due to the pandemic, events for using the digital patch were organised deliberately with this study in mind and data outside of in crowd events and with non-wearers, is therefore scarce. We hope this can be studied in more detail, once limitations due the ongoing pandemic are eased. 

%long term study ->The value of a badge comes from the interactions around, e.g. achieving a high amount of points in a pub crawl. through this patches are meaningful elements in their adornment practices. Similarly, Digi Merkki and its pictures become meaningful through the histroy attached to it from trading with other students. 

Finally, Digi Merkki’s function as a social lubricant was likely reinforced via its relative novelty. A continuous influx of new students and consequently new pictures would probably incentivise continuous use. 
%, the topic of long-term adoption of social wearables is beyond the scope of this study.
In the case of boiler suits, Digi Merkki’s function would likely shift from adornment to reminiscing, once students leave their university life behind. Nevertheless, more longitudinal studies on the practices around wearing technology that augments social interactions are necessary to understand what shapes adoption (and abandonment) of social wearables over time.

\section{Conclusion}
We study a social wearable in a distinct socio-cultural context to identify guiding design principles for the adoption of social wearables.
Our design proved to be effective in augmenting the participants’ existing social practices. Additionally, we reported on how our \study{intervention} generated emergent social practices, and participants navigated tensions in divergent social strategies through the help of memes. From this finding, we formulated the strong concept of “Memetic Expression” to guide future design.
When HCI aims to augment our co-present social interactions, scholars often focus on mediating specific interactions, such as digitising social signalling. If we design our technologies with an understanding that reduces communication to finite states (e.g. binary: available for engagement vs avoiding it), then our designs will be limited as well. In contrast, a perspective that respects the complexities of our social life needs to be open for participatory culture as showcased by internet memes.
% By designing social wearables we bring digital communication technology into people’s clothing. As observed in our study this change makes clothing part of the networked society. Networked society can be described as a participatory culture, full of imitation and bottom up remix of content and meanings. Hence, it is plausible our clothing will ultimately become part of remix culture and designers can make us of this fact. 

\begin{acks}
We want to thank our co-designers Rasmus Ruhola, Anna He, Tapio “Tassu” Takala, Ilyena Hirskyj-Douglas, Emmi Puota and Maria Serena Ciaburri, as well as Antti Salovaara for his guidance. This work was supported by the Academy of Finland project grant Digital Aura (311090). Additionally the work by Anna Kantasalo was supported by CACDAR (328729) and Felix A. Epp by Future Methods (330124).
\end{acks}

\bibliographystyle{ACM-Reference-Format}
\bibliography{fixes,adorned-in-memes}

% \appendix



\end{document}